\documentclass[%
 aip,
 amsmath,amssymb,
 reprint,%
]{revtex4-1}

\usepackage{graphicx}
\usepackage{dcolumn}
\usepackage{bm}

\usepackage[utf8]{inputenc}
\usepackage[T1]{fontenc}
\usepackage{mathptmx}
\usepackage{etoolbox}

\makeatletter
\def\@email#1#2{%
 \endgroup
 \patchcmd{\titleblock@produce}
  {\frontmatter@RRAPformat}
  {\frontmatter@RRAPformat{\produce@RRAP{*#1\href{mailto:#2}{#2}}}\frontmatter@RRAPformat}
  {}{}
}%
\makeatother
\begin{document}

\preprint{AIP/123-QED}

\title[Optimal Demodulation Domain for Microwave SQUID
Multiplexers in Presence of Readout System Noise.]{Optimal Demodulation Domain for Microwave SQUID
Multiplexers in Presence of Readout System Noise.}
\author{M. E. García Redondo}
    \affiliation{Instituto de Tecnologías en Detección y Astropartículas (ITeDA), Buenos Aires, Argentina}
    \affiliation{Institute for Data Processing and Electronics (IPE), Karlsruhe Institute of Technology (KIT), Karlsruhe, Germany}
    \affiliation{Comisión Nacional de Energía Atómica (CNEA), Buenos Aires, Argentina}
    \affiliation{Universidad Nacional de San Martín (UNSAM), Buenos Aires, Argentina}
    \email{manuel.garcia@iteda.cnea.gov.ar.}
\author{N. A. Müller}
    \affiliation{Instituto de Tecnologías en Detección y Astropartículas (ITeDA), Buenos Aires, Argentina}
    \affiliation{Institute of Micro- and Nanoelectronic Systems (IMS), Karlsruhe Institute of Technology (KIT), Karlsruhe, Germany}
    \affiliation{Comisión Nacional de Energía Atómica (CNEA), Buenos Aires, Argentina}
    \affiliation{Universidad Nacional de San Martín (UNSAM), Buenos Aires, Argentina}
\author{J. M. Salum}
    \affiliation{Instituto de Tecnologías en Detección y Astropartículas (ITeDA), Buenos Aires, Argentina}
    \affiliation{Institute for Data Processing and Electronics (IPE), Karlsruhe Institute of Technology (KIT), Karlsruhe, Germany}
    \affiliation{Comisión Nacional de Energía Atómica (CNEA), Buenos Aires, Argentina}
    \affiliation{Universidad Nacional de San Martín (UNSAM), Buenos Aires, Argentina}
\author{L. P. Ferreyro}
    \affiliation{Instituto de Tecnologías en Detección y Astropartículas (ITeDA), Buenos Aires, Argentina}
    \affiliation{Institute for Data Processing and Electronics (IPE), Karlsruhe Institute of Technology (KIT), Karlsruhe, Germany}
    \affiliation{Consejo Nacional de Investigaciones Científicas y Técnicas (CONICET), Buenos Aires, Argentina}
    \affiliation{Universidad Nacional de San Martín (UNSAM), Buenos Aires, Argentina}
\author{J. D. Bonilla-Neira}
    \affiliation{Instituto de Tecnologías en Detección y Astropartículas (ITeDA), Buenos Aires, Argentina}
    \affiliation{Institute of Micro- and Nanoelectronic Systems (IMS), Karlsruhe Institute of Technology (KIT), Karlsruhe, Germany}
    \affiliation{Consejo Nacional de Investigaciones Científicas y Técnicas (CONICET), Buenos Aires, Argentina}
    \affiliation{Universidad Nacional de San Martín (UNSAM), Buenos Aires, Argentina}
\author{J. M. Geria}
    \affiliation{Instituto de Tecnologías en Detección y Astropartículas (ITeDA), Buenos Aires, Argentina}
    \affiliation{Institute of Micro- and Nanoelectronic Systems (IMS), Karlsruhe Institute of Technology (KIT), Karlsruhe, Germany}
    \affiliation{Consejo Nacional de Investigaciones Científicas y Técnicas (CONICET), Buenos Aires, Argentina}
\author{J. J. Bonaparte}
    \affiliation{Comisión Nacional de Energía Atómica (CNEA), Buenos Aires, Argentina}
    \affiliation{Institute of Micro- and Nanoelectronic Systems (IMS), Karlsruhe Institute of Technology (KIT), Karlsruhe, Germany}
    \affiliation{Universidad Nacional de San Martín (UNSAM), Buenos Aires, Argentina}
\author{T. Muscheid}
    \affiliation{Institute for Data Processing and Electronics (IPE), Karlsruhe Institute of Technology (KIT), Karlsruhe, Germany}
\author{R. Gartmann}
    \affiliation{Institute for Data Processing and Electronics (IPE), Karlsruhe Institute of Technology (KIT), Karlsruhe, Germany}

\author{A. Almela}
    \affiliation{Instituto de Tecnologías en Detección y Astropartículas (ITeDA), Buenos Aires, Argentina}
    \affiliation{Comisión Nacional de Energía Atómica (CNEA), Buenos Aires, Argentina}
\author{M. R. Hampel}
    \affiliation{Instituto de Tecnologías en Detección y Astropartículas (ITeDA), Buenos Aires, Argentina}
    \affiliation{Comisión Nacional de Energía Atómica (CNEA), Buenos Aires, Argentina}
    \affiliation{Consejo Nacional de Investigaciones Científicas y Técnicas (CONICET), Buenos Aires, Argentina}
\author{A. E. Fuster}
    \affiliation{Instituto de Tecnologías en Detección y Astropartículas (ITeDA), Buenos Aires, Argentina}
    \affiliation{Comisión Nacional de Energía Atómica (CNEA), Buenos Aires, Argentina}
    \affiliation{Consejo Nacional de Investigaciones Científicas y Técnicas (CONICET), Buenos Aires, Argentina}
\author{L. E. Ardila-Perez}
    \affiliation{Institute for Data Processing and Electronics (IPE), Karlsruhe Institute of Technology (KIT), Karlsruhe, Germany}
\author{M. Wegner}
    \affiliation{Institute for Data Processing and Electronics (IPE), Karlsruhe Institute of Technology (KIT), Karlsruhe, Germany}
    \affiliation{Institute of Micro- and Nanoelectronic Systems (IMS), Karlsruhe Institute of Technology (KIT), Karlsruhe, Germany}

\author{M. Platino}
    \affiliation{Instituto de Tecnologías en Detección y Astropartículas (ITeDA), Buenos Aires, Argentina}
    \affiliation{Comisión Nacional de Energía Atómica (CNEA), Buenos Aires, Argentina}
    \affiliation{Consejo Nacional de Investigaciones Científicas y Técnicas (CONICET), Buenos Aires, Argentina}
    \affiliation{Universidad Nacional de San Martín (UNSAM), Buenos Aires, Argentina}
\author{O. Sander}
    \affiliation{Institute for Data Processing and Electronics (IPE), Karlsruhe Institute of Technology (KIT), Karlsruhe, Germany}
\author{S. Kempf}
    \affiliation{Institute of Micro- and Nanoelectronic Systems (IMS), Karlsruhe Institute of Technology (KIT), Karlsruhe, Germany}
\author{M. Weber}
    \affiliation{Institute for Data Processing and Electronics (IPE), Karlsruhe Institute of Technology (KIT), Karlsruhe, Germany}

\date{\today}

\begin{abstract}
The Microwave SQUID Multiplexer (\textmu MUX) is the device of choice for the readout of a large number of Low-Temperature Detectors in a wide variety of experiments within the fields of astronomy and particle physics. While it offers large multiplexing factors, the system noise performance is highly dependent on the cold and warm-readout electronic systems used to read it out, as well as the demodulation domain and parameters chosen. In order to understand the impact of the readout systems in the overall detection system noise performance, first we extended the available \textmu MUX simulation frameworks including additive and multiplicative noise sources in the probing tones (i.e. phase and amplitude noise), along with the capability of demodulating the scientific data, either in resonator's phase or scattering amplitude. Then, considering the additive noise as a dominant noise source, the optimum readout parameters to achieve minimum system noise were found for both open-loop and flux-ramp demodulation schemes in the aforementioned domains. Later, we evaluated the system noise sensitivity to multiplicative noise sources under the optimum readout parameters. Finally, as a case study, we evaluated the optimal demodulation domain and expected system noise level for a typical Software-Defined Radio (SDR) readout system. This work leads to an improved system performance prediction and noise engineering based on the available readout electronics and selected demodulation domain.
\end{abstract}

\maketitle


\section{\label{sec:intro}Introduction\protect}

Low-Temperature Detectors such as Transition-Edge Sensors (TES)\cite{Walker2020}, Magnetic Microcalorimeters (MMC)\cite{Fleischmann2009} and Microwave Kinetic Inductance Detectors (MKID)\cite{Zmuidzinas2012} have demonstrated outstanding sensitivities in a large number of different experiments going from sub-mm/mm\cite{Piat2020} and gamma/X-ray\cite{Ullom2015} astronomy to nuclear and particle physics\cite{Gastaldo2017}. Nowadays, current micro and nano-fabrication techniques allow to easily create focal planes populated with a huge number of detectors satisfying the requirements of the most demanding applications\cite{Duff2016}. At the same time, multiplexing schemes that support these large detector counts are being developed in parallel in order to manage the system complexity and cooling requirements at sub-Kelvin stages where the detectors are operated\cite{Abitbol2017}.

Along with the different multiplexing techniques developed for cryogenic detectors readout, a Frequency-Division Multiplexing (FDM) scheme that has become popular in the last decade is the Microwave Superconducting Quantum Interference Device (SQUID) Multiplexing (\textmu MUXing)\cite{Irwin2004,matesthesis}. Its success lies in the available high channel carrying capacity and remarkable system scalability while maintaining the readout noise subdominant to the intrinsic detector noise\cite{Dober2021}. This scheme encodes the detector signals in the resonance frequencies of multiple GHz-frequency superconducting resonators coupled to a common feedline. Therefore, recovering the signals from each detector requires only monitoring the resonance frequencies. Despite these advantages, it places stringent requirements on the cold and warm readout systems responsible for generating and acquiring high-purity broadband microwave signals required to monitoring each channel.

Particularly for low-detector count experiments with wider bandwidth resonators (e.g. as used for X-ray pulse detection\cite{Kempf2017}) the readout system noise is mainly dominated by the cryogenic Low-Noise Amplifier (LNA) in charge of recovering the low-power microwave signals coming out of the \textmu MUX device\cite{lynxread}. Contrary, for high-detector count experiments using narrower resonances (e.g. as used for CMB surveys\cite{Dober2021}) readout system noise tends to be dominated by the Two-Level System noise (TLS)\cite{matesthesis}. For the case of single channel in-lab characterization, the conditions in which the system noise is dominated by cryogenic LNA or TLS noise are easily achieved using bench-top instruments and expensive microwave components. This is not the case for in-situ measurements in either ground-based or space-based experiments. These kinds of experiments ussually use custom designed Software-Defined Radio (SDR) readout systems based on commercially available components with limited performance in comparison with bench-top instruments\cite{Gartmann2022,Yu2023,Redondo2023rfsoc,lynxrfsoc}. Therefore, the impact of the readout electronics becomes more relevant, especially in the readout of flux-transformer coupled devices such as Magnetic Microcalorimeters/bolometers\cite{Kempf2017,Geria2022} or with the introduction of near–quantum-limited amplification technologies such as parametric amplifiers\cite{Zobrist19,Malnou}. As a consequence, it is important to have a software simulation framework which allows designers to predict the impact of these systems on the readout performance, as well as, through its characterization, evaluate the readout parameters that optimize the yield of a particular detection system.

Seeking to improve system noise predictions and find optimization parameters, we extended in this article the capabilities of the available simulation frameworks for the readout of Microwave SQUID Multiplexers\cite{yusim,Schuster2023} allowing to evaluate the impact of the cold and warm readout electronics systems in the overall readout performance. Previously developed frameworks solely consider additive noise, commonly attributed to the cryogenic LNA. However, characterizations preformed over SDR readout systems unveil the presence of multiplicative noise\cite{Rantwijk,Redondo2023rfsoc,Herrmann22,exp_quick}. These extra noise sources were included in the present study and their impact in the system performance was evaluated as a function of the readout parameters. Since the noise sources added act in different domains, the most commonly used demodulation domains were also included. In order to avoid confusing readers, throughout this document we will use the term ``domain'' to refer to the coordinate system where the signal from a single detector is encoded. This term should not be confused with the ``space'' where the different detectors are multiplexed (e.g. time, frequency or code). For the case of \textmu MUX, which is a multiplexing device in the frequency space, we will analyze the two most commonly used readout domains according to the available literature. As we will detail later, the two demodulation domains are the amplitude of the transmission scattering parameter $\gamma$ and the resonator's phase $\theta$. It is important to mention that the scope of this work only considers the readout by a single fixed tone and is not directly applicable to the analysis of other recently proposed readout techniques such as Tone-tracking~\cite{Yu2023,yusim}.

This article is organized as follows: First, in section \ref{sec:umuxmodel} we describe the \textmu MUX model used for simulations and its dependence with flux and power. Second, section \ref{sec:readout} presents the readout signal model adding additional noise sources attributed to the readout system and two demodulation schemes and domains are presented. Then, in section \ref{sec:sysnoise} we show through simulations the minimum-noise readout parameters for additive noise in both demodulation schemes and evaluate the impact of the remaining noise sources under these conditions using a variety of different demodulation parameters. In addition, the impact of a typical SDR readout system is analyzed as an example. Finally, a discussion based on the obtained results is given in section \ref{sec:disc} and the conclusions are presented in section \ref{sec:conc}.

\section{\label{sec:umuxmodel}Microwave SQUID Multiplexer Model}

Evaluating the readout performance of a \textmu MUX under readout system noise first requires a model capable of replicating the complex underlying physics of superconducting resonators combined with the nonlinear behaviour of Josephson tunnel junctions. In this section we introduce the most important aspects of \textmu MUX model that we will employ in section \ref{sec:readout}. It is worth to mention that the main objective of this work is not to improve the current \textmu MUX models, but rather to reuse the available ones\cite{Wegner_2022,Schuster2023} within a new simulation framework that includes new features and noise sources.

\begin{figure}
\includegraphics[width=0.48\textwidth]{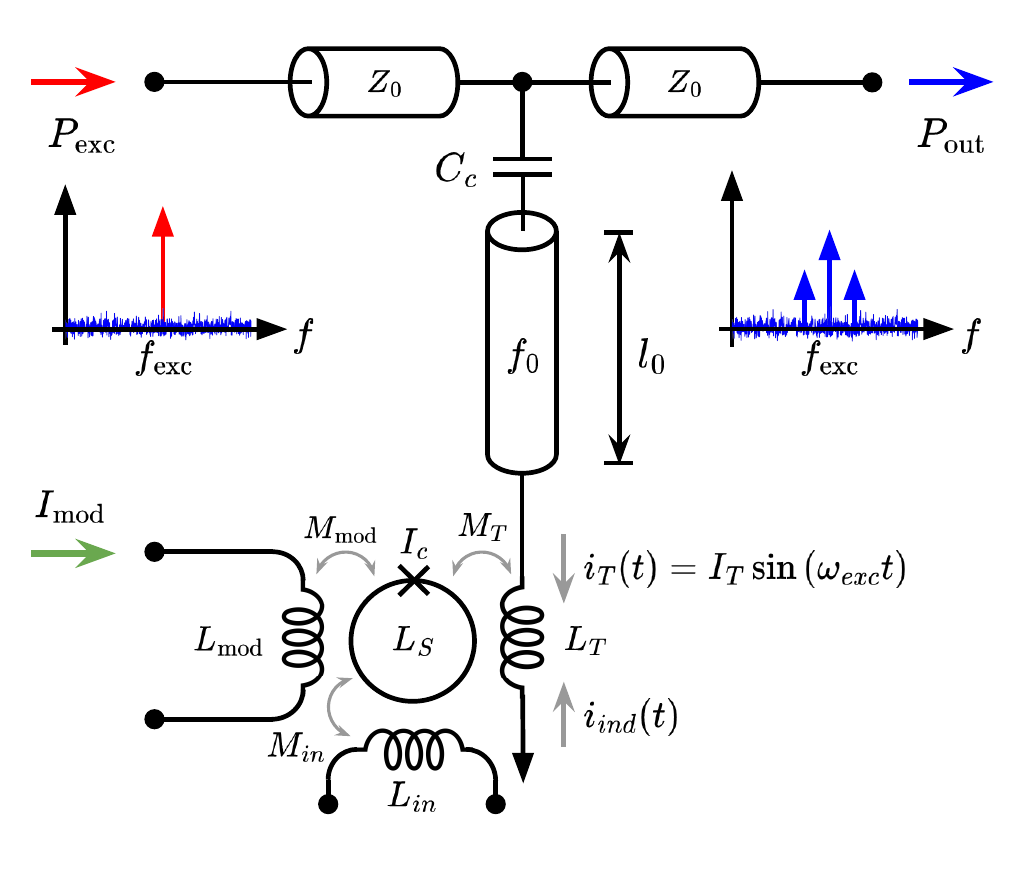}
\caption{\label{fig:umux} Simplified schematic circuit diagram of a single readout channel of a transmission line based Microwave SQUID Multiplexer. Left) Probing tone spectrum with frequency $f_{\mathrm{exc}}$ and power $P_{\mathrm{exc}}$. Right) Output tone spectrum with its modulation sidebands produced by the \textmu MUX.}
\end{figure}

\textmu MUXing relies on High-Q superconducting microwave resonators as frequency encoding elements. Figure \ref{fig:umux} shows a simplified schematic circuit diagram of a transmission line based \textmu MUX. It comprises of a non-hysteretic current-sensing rf-SQUID inductively coupled to a superconducting quarter-wave transmission line resonator. The resonator, in turn, is capacitively coupled to a common feedline. Due to its Josephson parametric self-inductance, the rf-SQUID acts as a flux-dependent inductor allowing the resonance frequency modulation by means of the flux threading the SQUID loop, which is proportional to the detector signal coupled through the detector input coil. As a consequence, a probing tone injected at the multiplexer input (Left spectrum in figure \ref{fig:umux}) will appear modulated in phase and amplitude at the multiplexer output (Right spectrum in figure \ref{fig:umux}). In this way, demodulating the output tones corresponding to each resonator allows us to decode the detector signals using a single pair of coaxial cables going in and out of the cryostat.

As in communications systems, noise performance in the presented scheme depends on the type of modulation used, signal power and noise level. Accordingly, we will start describing the modulation produced by the \textmu MUX in the probing tones. In the following subsection \ref{subsec:powerdep}, we will present the resonance frequency flux and power dependence. Then in subsection \ref{subsec:freqres}, we will relate it with the resonator frequency response, and finally in subsection \ref{subsec:example} introduce an example model that will be used in the next sections \ref{sec:readout} and \ref{sec:sysnoise}.

\subsection{\label{subsec:powerdep}Resonance Frequency Flux and Power Dependence}

For the purpose of this work, we directly adopted the analytical \textmu MUX readout power dependent model developed by Wegner\cite{Wegner_2022} to describe the dependence of the resonance frequency as a function of the fluxes contributions threading SQUID loop $f_r(\varphi)$. This model assumes a finite Self-screening parameter $\beta_L=2 \pi L_s I_c/\Phi_0$ as well as a non-zero readout available power $P_{\mathrm{exc}}$ at the \textmu MUX input port with angular frequency $\omega_{\mathrm{exc}}=2 \pi f_{\mathrm{exc}}$. Under this condition, the supercurrent in the rf-SQUID can be written as,

\begin{equation}
    I_S(t)=-I_c \sin{\left[\varphi_{\mathrm{mod}}+\varphi_{\mathrm{exc}}\sin{(\omega_{\mathrm{exc}}t)}+\beta_L\frac{I_{S}(t)}{I_c}\right]}
    \label{eq:supcur}
\end{equation}

Where $\varphi_{\mathrm{mod}}=2 \pi M_{\mathrm{mod}} I_{\mathrm{mod}}/\Phi_0$ is a quasi-static magnetic flux contribution induced by a current source $I_{\mathrm{mod}}$ coupled through $M_{\mathrm{mod}}$, $\varphi_{\mathrm{exc}}=2 \pi M_T I_{T}/\Phi_0$ is the amplitude of the time-variant magnetic flux induced by the probing tone through $M_T$ and $I_c$ the critical current of the Josephson junction. The term $\varphi_{\mathrm{scr}}(t)=\beta_L I_{S}(t)/I_c$ represents the self-screening flux. According to Lenz's law, a voltage $u_{\mathrm{ind}}(t)=-M_T d I_{\mathrm{s}}(t)/dt$ is created which opposes the flux changes and hence the current $i_{\mathrm{ind}(t)}$ is induced in the resonator termination. Therefore, the total current at the resonator termination $i_{\mathrm{tot}}(t)=i_T(t)+i_{\mathrm{ind}}(t)$ is a superposition of two contributions originating from the microwave probing tone and the supercurrent flowing through the SQUID loop. In this way, the total voltage across the resonator termination is $u_{\mathrm{ind}}(t)=L_T d i_{\mathrm{tot}}(t)/dt$ and can be expressed by introducing an effective termination inductance,

\begin{equation}
    L_{T,\mathrm{eff}}=L_{T}\frac{i_{\mathrm{tot}}(t)}{i_{T}(t)}
    \label{eq:leff}
\end{equation}

Due to the transcendental nature of equation \ref{eq:supcur}, there is no analytical solution for $I_S(t)$ than can be directly inserted into equation \ref{eq:leff}. The required approximations and expansions of $I_S(t)$ in order to obtain a useful expression for $L_{T,\mathrm{eff}}$ are described by Wegner\cite{Wegner_2022}. The derived expression can be directly related to the circuit parameters shown in figure \ref{fig:umux} yielding to the following resonance frequency expression as a function of the normalized fluxes $f_r(\varphi_{\mathrm{exc}},\varphi_{\mathrm{mod}})$,

\begin{equation}
    f_r(\varphi_{\mathrm{exc}},\varphi_{\mathrm{mod}}) \approx f_{\mathrm{off}}+\frac{4f_0^2 M_T^2}{Z_0 L_S}\frac{2\beta_L}{\varphi_{\mathrm{exc}}} \sum_{i,j} p_{i,j}(\varphi_{\mathrm{exc}},\varphi_{\mathrm{mod}})
    \label{eq:fr}
\end{equation}

Where $f_{\mathrm{off}}=f_0 - 4f_0^2(C_c Z_0+ L_T/Z_0)$ stands for the unaltered resonance frequency and $p_{i,j}=a_{i,j}\beta_L^{b_{i,j}}J_1(c_{i,j}\varphi_{exc})\cos{(c_{i,j}\varphi_{mod})}$ are the Taylor expansion coefficients. Here, $i$ denotes the expansion order and $j$ the number of contribution of each order, while $J_1$ is the order 1 Bessel function of the first kind. It is not the purpose of this article to derive the model equations in detail again, but to give the basic necessary elements to understand the following sections. $p_{i,j}$ coefficients, as well as a detailed analysis can be found in the following references \cite{Wegner_2022,ahrensthesis}. The validity of this model sets an upper limit for $\beta_L\leq 0.6$ and assumes a probing tone frequency $f_{exc}$ close to the resonance frequency of the resonator $f_{r}$, leading to a maximum oscillating anti-node current amplitude $I_T=\sqrt{\frac{16 Q_{l}^2P_{exc}}{\pi Q_c Z_0}}$. $Q_l$ and $Q_c$ denote the loaded and coupling quality factors respectively, while $Z_0$ is the characteristic impedance of both the resonator and the feedline.

\subsection{\label{subsec:freqres}Frequency Response}
The previously introduced expression can be inserted into the resonator transmission scattering parameter in order to find the flux dependence of the resonator frequency response $S_{21}(f_{\mathrm{exc}},\varphi)$. Considering a perfectly symmetric lossy resonator\cite{Probst,Schuster2023}, the analytical expression becomes,

\begin{equation}
    S_{21}(f_{\mathrm{exc}},\varphi_{\mathrm{exc}},\varphi_{\mathrm{mod}})=\frac{S_{21}^{\mathrm{min}}+2jQ_l\left(\frac{f_{\mathrm{exc}}-f_r(\varphi_{\mathrm{exc}},\varphi_{\mathrm{mod}})}{f_r(\varphi_{\mathrm{exc}},\varphi_{\mathrm{mod}})}\right)}{1+2jQ_l\left(\frac{f_{\mathrm{exc}}-f_r(\varphi_{\mathrm{exc}},\varphi_{\mathrm{mod}})}{f_r(\varphi_{\mathrm{exc}},\varphi_{\mathrm{mod}})}\right)}
    \label{eq:freqres}
\end{equation}

\begin{figure*}
\includegraphics[width=0.9\textwidth]{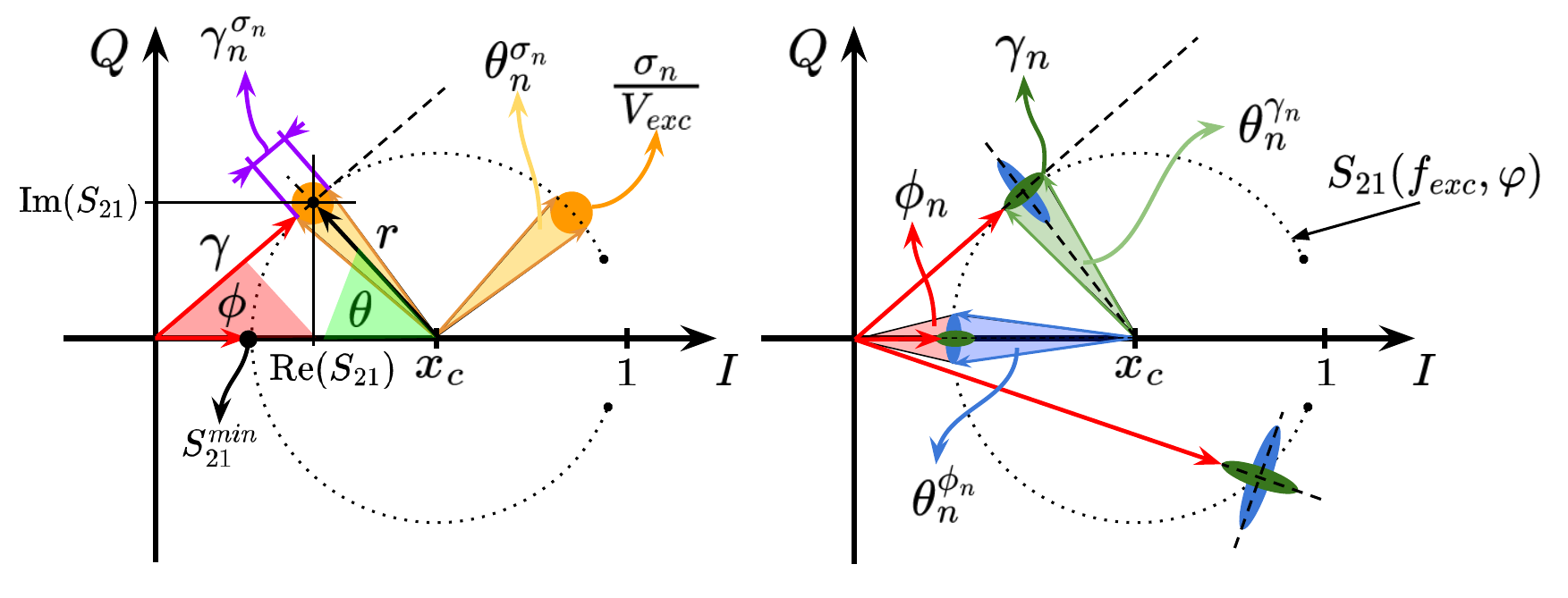}
\caption{\label{fig:trajectories} Complex plane representation of the transmission scattering parameter $S_{21}(f_{\mathrm{exc}},\varphi)$ in its Canonical form, with the readout noise sources. Red arrows represent the normalized measured signal phasor with amplitude $\gamma=|S_{21}|$ and phase $\phi$, while $\theta$ is the resonator's phase measured from its rotating frame. Left) Additive source: Johnson–Nyquist noise $\sigma_n(t)/V_{\mathrm{exc}}$ represented as a circular orange noise cloud. Right) Multiplicative sources: Phase $\phi_n(t)$ and Amplitude $\gamma_n(t)$ noises represented with the blue and green ellipses respectively. The projection of each noise source into the resonator's phase and scattering amplitude demodulation domains are denoted as: $\gamma_n^{\sigma_n}(t)$, $\theta_n^{\sigma_n}(t)$, $\theta_n^{\gamma_n}(t)$ and $\theta_n^{\phi_n}(t)$. The superscript stands for the noise source that is being projected.} 
\end{figure*}

Here $S_{21}^{\mathrm{min}} \approx Q_l/Q_i$ represents the resonance depth, $Q_l$ and $Q_i$ are the loaded and intrinsic quality factors respectively. This equation describes a Lorentzian shaped frequency response that in the complex plane follows a semi-circular trajectory with a center at $x_c=(1+S_{21}^{\mathrm{min}})/2$ and radius $r=1-x_c$ as can be seen in figure \ref{fig:trajectories}. This trajectory is approximately the same either for flux or frequency sweeps in the case of a negligible junction sub-gap conductance\cite{Kempf2017,ahrensthesis}. In the case of high quality resonators ($Q_i \rightarrow \infty$), the bandwidth of the resonator can be easily approximated using $BW_{\mathrm{res}}\approx f_0/Q_l$. Therefore, $Q_l \approx Q_c$ and $BW_{\mathrm{res}}$ is only determined via coupling through $C_c$. The symmetric frequency response described by equation \ref{eq:freqres} is a consequence of a good impedance matching between the resonator and the associated feed-line\cite{ahrensthesis,Probst}.

\subsection{\label{subsec:simmodel}\textmu MUX Simulation Example}

The \textmu MUX analytical model described by equations \ref{eq:fr} and \ref{eq:freqres} was numerically implemented using Python and tested with device parameters described in table \ref{tab:umuxparams}, that are typical for a \textmu MUX device used in bolometric applications\cite{Dober2021}. Using equation \ref{eq:fr} the unaltered resonance frequency was calculated yielding $f_{\mathrm{off}}=4.775$~GHz, while the peak-to-peak frequency shift $\Delta f_r^{\mathrm{mod}}$ for vanishing readout powers $P_{\mathrm{exc}} \rightarrow 0$ matches the resonator bandwidth $BW_{\mathrm{res}}\approx 200$~kHz. Since we commonly place the probing tone frequency $f_{\mathrm{exc}}$ close to the unaltered resonant frequency $f_{\mathrm{off}}$, it is more convenient to express it in terms of the difference $f_{\mathrm{exc}}-f_{\mathrm{off}}$ instead of its absolute frequency. We adopted this metric for our graphs in this work. Accordingly to the analytical model predictions at a power around $P_{\mathrm{exc}} \approx -64$~dBm the resonator became insensitive to the modulation flux $\varphi_{\mathrm{mod}}$. For this reason, simulations performed in this work were limited to a power range between $-90$~dBm and $-60$~dBm.

\begin{table}
\caption{\label{tab:umuxparams} \textmu MUX example parameters used during simulations. These are the typical design parameters for a \textmu MUX optimized for bolometric applications\cite{Dober2021}.}
\begin{ruledtabular}
\begin{tabular}{lcr}
Parameter&Value&Unit\\
\hline
$f_{0}$& 5 &  GHz\\
$Z_{0}$& 50  & $\Omega$\\
$L_{S}$& 30 &  pH\\
$L_{T}$& 100 &  pH\\
$\beta_{L}$ & 0.6 & -\\
$M_{T}$& 1.3 & pH\\
$M_{\mathrm{mod}}$& 20 & pH\\
$Q_{i}$ & 200000 & -\\
$C_{c}\footnote{$Q_c \approx (2 \pi)/(2 Z_0 \omega_0 C_c)^2$}$ & 5 & fF\\
\end{tabular}
\end{ruledtabular}
\end{table}

It should be noted that the required condition of a probe tone frequency $f_{\mathrm{exc}}$ being as close as possible to the resonance frequency $f_{r}$ cannot be always satisfied, yielding to an overestimation of the radio-frequency flux induced within the SQUID when the probing tone is far away from it. In this case, an analytical calculation of $I_T(f_{\mathrm{exc}})$ cannot be found due to the recursive relation between $f_r(\varphi_{\mathrm{exc}})$ and $\varphi_{\mathrm{exc}}(f_r)$. Therefore, we applied a similar solution that was used in previous articles\cite{Schuster2023}. But in this case, the maximum value of the anti-node current $I_T=\sqrt{\frac{16 Q_{l}^2P_{\mathrm{exc}}}{\pi Q_c Z_0}}$ was scaled using narrow-band approximation of the transmission line resonator frequency response given by expression \ref{eq:freqres}. This iterative process starts at a zero readout power condition $P_{\mathrm{exc}} \rightarrow 0$ to derive a first guess $f_r^{i=0}$ and then executes the power dependent model sequentially using the previous results $f_r^{i-1}$ until the equilibrium is reached or equivalently the convergence criteria is satisfied $f_r^{i-1}\approx f_r^{i}+\epsilon$.

Due to the fact that in the next section we use a continuous-time analytical description, time-dependency was introduced considering a time varying modulation flux $\varphi_{\mathrm{mod}}(t)$. The quasi-static flux condition imposed for equation \ref{eq:supcur} is fulfilled ensuring flux variations considerably slower than the resonator ring-down time $\tau_{\mathrm{down}}\approx 1/2 \pi BW_{\mathrm{res}}$ and consequently even slower than $\tau_{\mathrm{exc}}=1/2 \pi f_{\mathrm{exc}}$. This condition allowed the resonator to keep track of the flux variations while maintaining the steady-state frequency response\cite{Schuster2023,Mates2012} described by equation \ref{eq:freqres}. Nevertheless, our model allows us to calculate a first order approximation of the dynamic response by means of a low-pass filtered version of $S_{21}$. In order to simplify the nomenclature and facilitate the following analysis, we will only show explicitly the time dependence of the transmission parameter $S_{21}(t)$ as a consequence of the flux variations $\varphi(t)$.

\section{\label{sec:readout}Readout System Model}

\begin{figure*}
\includegraphics[width=0.95\textwidth]{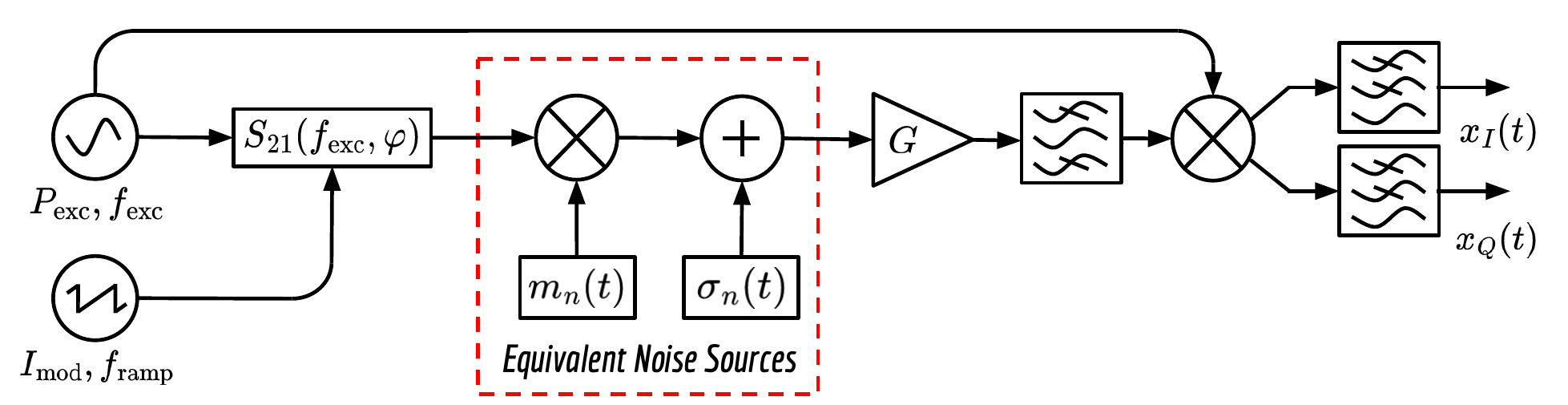}
\caption{\label{fig:readout} Block diagram of the homodyne readout system used for the readout of a Microwave SQUID Multiplexer. From left to right) Microwave and arbitrary signal generators in charge of generating the probing tone and flux modulation, \textmu MUX device, equivalent noise sources, low-noise amplifiers, filters and IQ mixer for signal conditioning and down-conversion. The noise sources inside the red rectangle $\sigma_n(t)$, $m_{n}(t)$ represent the additive and multiplicative noise sources attributed to the cold and warm readout system respectively and referred to the \textmu MUX output.}
\end{figure*}

A simplified block diagram of the homodyne system used for the readout of a single \textmu MUX channel is shown in figure \ref{fig:readout}. For the sake of simplicity, several of the microwave/RF components comprising the cold and the warm-temperature electronics were condensed into single blocks while preserving its functionality. The most important components of this system are: 1) A microwave synthesizer capable of generating the probing tone with frequency $f_{\mathrm{exc}}$, and available power $P_{\mathrm{exc}}$ at the \textmu MUX input port, 2) An arbitrary current waveform generator able to provide the required flux modulation $\varphi_{\mathrm{mod}}$ with frequency $f_{\mathrm{ramp}}$, 3) The \textmu MUX device described by the behavioural model explained in section \ref{sec:umuxmodel}, 4) A set of noise sources representing all noise contributions attributed to the readout system (dashed red box), 5) Low-Noise amplification and filtering stages in charge of boosting the signals and filtering out unwanted components maintaining the Signal-to-Noise Ratio (SNR) and, 6) A complex IQ mixer that down-converts the signal in order to be processed later.

\subsection{\label{subsec:signalmeas}Measured Signals and Noise Sources}

The complex base-band signal $x(t)=x_{I}(t)+jx_{Q}(t)$ at the output of the homodyne readout system shown in figure \ref{fig:readout} can be expressed as,

\begin{equation}
    x(t)=G^{\prime} V_{\mathrm{exc}} S_{21}(t)m_n(t)+G^{\prime} \sigma_n(t)
    \label{eq:signalmeas}
\end{equation}

Where $V_{\mathrm{exc}}=\sqrt{2P_{exc}Z_0}$ is the amplitude of the probing tone, while $\sigma_n(t)$ and $m_n(t)$ are the additive and multiplicative noise terms respectively. Here, we consider an ideal IQ mixer and a gain factor $G^{\prime}=G e^{j\alpha}e^{2j \pi f_{\mathrm{exc}} \tau}$ including the overall gain $G$, phase offset $\alpha$ and time delay $\tau$ added by the RF components and cables. In the context of a real measurement, $G^{\prime}$ can be determined during the calibration process or continuously monitored \cite{Silva2022}. Therefore, for the purpose of this article we will assume that all of the aforementioned parameters are known and constant over time for the probing tone frequency $f_{\mathrm{exc}}$. During the down-conversion process the time dependence of the excitation signal was removed keeping only the complex envelope of the modulated probing tone being equal to the scaled transmission parameter time-trace $G^{\prime}S_{21}(t)$. When $f_{\mathrm{exc}}$ is located out of resonance, the transmission parameter is close to unity and the measured average signal amplitude approximately equals to $x^{\mathrm{off}}(t) \approx G^{\prime}V_{\mathrm{exc}}$. This value can be used to define a normalized signal as,

\begin{equation}
    \frac{x(t)}{x^{\mathrm{off}}(t)}=S_{21}^{\mathrm{meas}}(t)=S_{21}(t)m_n(t)+\frac{\sigma_n(t)}{V_{\mathrm{exc}}}
    \label{eq:meassig}
\end{equation}

Thus, the normalized signal is basically an estimation of the actual transmission parameter $S_{21}(t)$. Figure \ref{fig:trajectories} shows the complex representation of $S_{21}^{\mathrm{meas}}$ where, as we mention in section \ref{subsec:freqres} and for the noise-less case, the measured signal (red arrow) describes a semi-circle centered at $x_c$ with radius $r$ as a consequence of a flux sweep. The $|S_{21}^{\mathrm{meas}}|$ takes values from $S_{21}^{min}$ on resonance, to $S_{21} \approx 1$ far out of resonance. Contrarily, in a real scenario represented by equation \ref{eq:signalmeas}, noise sources produce a deviation from this trajectory that cannot be distinguished form the flux variations produced by the detector signal and are interpreted as a flux noise. In the next subsections we give a detailed description of the noise sources affecting the system in order to calculate the error in the transmission parameter determination and consequently derive the equivalent flux noise represented by each one. Since we want to analyze the impact of the readout system noise over different demodulation domains, we did not consider detector, SQUID and modulation flux noise sources.

\subsubsection{\label{subsec:add}Additive Sources}

This type of noise refer to all kinds of noise sources that are directly added to the desired signals and remain even if the desired signal is not present. Considering the scope of this work, and based on previous articles\cite{Schuster2023,Malnou}, we will only consider Johnson–Nyquist noise $\sigma_n(t)$. Johnson–Nyquist noise can be modeled as $\sigma_n(t)=\sigma_I(t)+j \sigma_Q(t)$, where real and imaginary components are both zero-mean, finite power, Independent and Identically Distributed (IID) Gaussian random variables. A normalized version $\sigma_n(t)/V_{\mathrm{exc}}$ can be seen as an orange circular noise cloud in figure \ref{fig:trajectories}. This model is well suited to describe a wide range of noise sources present in readout electronics systems such as thermal, electric or quantization. In a properly designed system, the additive noise is typically dominated by the cryogenic High-Electron Mobility Transistor (HEMT) amplifier\cite{Rao2020}. Because of their importance, many groups are actively trying to improve them or even replace them with new amplification technologies such as quantum-limited parametric amplifiers\cite{Zobrist19,Malnou}. While these new technologies represent an important reduction of the additive noise level, they make evident the presence of other types of sources such as those we describe below. Additionally to the Johnson–Nyquist noise, spurious signals generated by the electronics\cite{Redondo2023rfsoc} can be also included as additive interferences, but for the scope of this work they will not be included because unlike noise, they are localized in frequency and can be separately mitigated\cite{Salum-espectral}.

\subsubsection{\label{subsec:mult}Multiplicative Sources}

Contrary to additive sources, multiplicative noise sources depend on the presence of the desired signal. Their level depend on the signal strength and mathematically, as the name implies, the noise is multiplied to the desired signal $S_{21}(t)$ as,

\begin{equation}
    m_n(t)=[1+\gamma_{n}(t)]e^{j\phi_n(t)}
    \label{eq:multnoise}
\end{equation}

Here, we will only focus on amplitude $\gamma_{n}(t)$ and phase $\phi_n(t)$ noise.  These are the most common multiplicative noise sources present in readout electronics systems\cite{Rubiola_2008,rohdemeas}. Both are real random variables described by their auto-correlation function as we will explain in subsection \ref{subsec:nsd}. A simplified and magnified representation of these noises using colored ellipses is shown in the right side of figure \ref{fig:trajectories}. The term $\gamma_{n}(t)$ represents small signal variations (green ellipses) parallel to the measured signal amplitude (red arrow), while $\phi_n(t)$ as perpendicular signal variations (blue ellipses). Due to their multiplicative nature, when the measured signal amplitude decreases (i.e., at resonance), the voltage fluctuations decrease proportionally keeping the amplitude and phase variations constant. Traditionally, amplitude and phase noise are terms used to describe short-term variations or instabilities, with ``short-term'' referring to time intervals on the order of seconds or less\cite{rohdepn}. These are commonly generated when some system parameter randomly fluctuates (e.g. due to thermal or flicker noise) translating that variation to the desired signal. This process called parametric up-conversion does not differ to much from the \textmu MUX working principle where the resonance frequency (system parameter) is modulated proportionally to the detector signal. This is the reason why the interest of this work is focused on including these noise sources and evaluating their impact on the system noise.

Besides being focused on noise sources generated by the readout system, this manuscript additionally includes the Two-Level System Noise. The evidence suggests that this noise is caused by Fluctuating Two-Level Systems (TLS) in dielectric materials, either the bulk substrate or its exposed surface, the interface layers between the metal films and the substrate, or any oxide layers on the metal surfaces that comprises the transmission line resonator\cite{Gao2007,Kumar}. This noise affects the distributed capacitance and therefore produces fluctuations in the resonance frequency that are seen as rotations of the resonance circle around its center. As well as amplitude noise in equation \ref{eq:multnoise}, it can be expressed as frequency fluctuations $f_r \cdot f_{\mathrm{TLS}}(t)$ around the unaltered resonance frequency $f_{\mathrm{off}}$.

\begin{equation}
    f_{\mathrm{r}}(t)=f_{\mathrm{r}}[1+f_{\mathrm{TLS}}(t)]
\end{equation}

Although TLS noise is not generated by the readout electronics, it can act differently depending on the demodulation domain and readout parameters as well as the readout noise sources. This is why its impact will be also evaluated. 

\subsection{\label{subsec:nsd}Noise Metrics}

Due to its random nature, noise is analyzed as a stochastic process. For wide-sense-stationary and ergodic random process $y(t)$, the Wiener-Khinchin theorem says that a power spectral density can be defined in terms of the Fourier transform of the statistical expected value, e.g. the auto-correlation\cite{Rubiola_2008} function as follows,

\begin{equation}
    S_{y}(\Delta f)=\mathcal{F} \left\{ R_{yy}(\tau)\right\}=\mathcal{F} \left\{ \left\langle y(0),y(\tau)\right\rangle \right\}
    \label{eq:nsd}
\end{equation}

Thus, spectral features of noise are entirely determined by the Noise Spectral Density $S_{y}(\Delta f)$ (NSD). The requirement that noise be stationary and ergodic is the equivalent of ``repeatable'' and ``reproducible'' in experimental physics. In our case, $S_{y}(\Delta f)$ is the base-band representation of the NSD measured at a frequency offset $\Delta f$ from the probing tone frequency $f_{\mathrm{exc}}$. For the case of Johnson–Nyquist noise, the statistic independence implies a white noise with constant NSD equals to $S_{N}=k_B T_{\mathrm{n}}$, where $T_{\mathrm{n}}$ is the so called noise temperature and $k_B$ the Boltzmann's constant. Therefore, the NSD of the normalized Johnson–Nyquist noise respect to the probing tone amplitude $\sigma_n(t)/V_{\mathrm{exc}}$ is,

\begin{equation}
    S_{\sigma}(\Delta f)=\frac{k_b T_{\mathrm{n}}}{P_{\mathrm{exc}}}
    \label{eq:addnsd}
\end{equation}

This quantity is usually expressed in dBc/Hz and $T_\mathrm{n}$ is the noise equivalent temperature referred to the \textmu MUX output. In the context of this work, we will distinguish between $T_\mathrm{n}$ and $T_{\mathrm{sys}}$ due to the fact that $T_{\mathrm{sys}}$ is commonly experimentally determined and as we will see later, corresponds to an equivalent noise temperature that takes into account all the different noise sources acting simultaneously and producing the same system noise level\cite{Malnou}.

On the other hand, amplitude, phase or TLS noise exhibits certain degree of correlation between realizations yielding to colored noise spectral densities. They are commonly described by a power law such as,

\begin{equation}
    S_{y}(\Delta f)=\sum_{n} b_n (\Delta f)^n
    \label{eq:powerlaw}
\end{equation}

With values of $n=[-4,0]$ for phase, $n=[-2,0]$ for amplitude\cite{Rubiola_2008} and $n=[-1,-1/2]$ for TLS\cite{Gao2007}. As in the case of Johnson–Nyquist, phase and amplitude noise are expressed in dBc/Hz and represent the noise power integrated in a $1$-Hz bandwidth at a $\Delta f$ offset from $f_{\mathrm{exc}}$ relative to the probe tone power. The TLS is expressed as fractional frequency fluctuations $S_{f_{\mathrm{r}}}(f)/f_{\mathrm{r}}^2$ in units of Hz$^{-1}$. Contrary to uncorrelated sources case, depending on the degree of correlation, noise can be removed. This is the case of phase noise in radio systems which use different oscillators and clocks locked to the same frequency reference during up-and down-conversion processes as well as during generation and sampling. Therefore, $\phi_n(t)$ used in equation \ref{eq:multnoise} stands for the so-called ``residual'' phase noise\cite{rohdemeas}. Due to the fact that it is a multiplicative quantity, it can be arbitrarily referred to any part of the circuit yielding the same behaviour.

\subsection{\label{subsec:example}Readout System Example}

\begin{figure}
\includegraphics[width=0.48\textwidth]{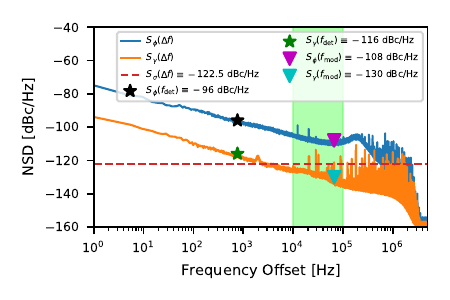}
\caption{\label{fig:readout-noise} Noise spectral densities for typical cold and warm-temperature readout systems at a frequency of $f_{\mathrm{exc}}=5$~GHz. Blue and orange solid lines are the residual phase and amplitude noise spectral densities. Dashed red line is the additive NSD calculated with equation \ref{eq:addnsd} using an equivalent noise temperature of a $T_{\mathrm{n}}=4$~K, and probing power $P_{\mathrm{exc}}=-70$~dBm. Black and green stars represent phase and amplitude noise measured at $f_{\mathrm{det}}\approx762$~Hz. The green region covers the possible modulation frequencies $f_{\mathrm{mod}}$ where the signal of interest is typically located in the case of bolometric applications\cite{Redondo2023rfsoc}. Particularly, we used $f_{\mathrm{mod}}\approx62$~kHz, where phase and amplitude noise take values represented by magenta and cyan triangles respectively.}
\end{figure}

As an example, figure \ref{fig:readout-noise} shows residual amplitude and phase noise spectral densities for a Direct-RF Software-Defined Radio (SDR) readout system based on the RFSoC ZCU216 evaluation kit\cite{zcu216}. These measurements were taken at a frequency of $f_{\mathrm{exc}}=5$~GHz using a loop-back cable connecting transmitter (Tx) and receiver (Rx) ensuring that most of the coherent noise is removed in the down-conversion/sampling process\cite{Redondo2023rfsoc}. In a single measurement at constant power, additive and multiplicative noises cannot be distinguished since both contribute to phase and amplitude noise. In order to separate the two contributions, a power sweep was performed until the amplitude and phase spectra reached power-independent values consistent with the description of the multiplicative noise. A more detailed description of the noise characterization process and measurement set-up is given in appendix~\ref{app:readnoisecharact}. The obtained results show that for frequency offsets below $10$~kHz amplitude $S_{\gamma}(\Delta f)$ and phase $S_{\phi}(\Delta f)$ noise densities have $-10$~dB/decade slopes demonstrating the colored behaviour described by equation \ref{eq:powerlaw}. In the case of amplitude noise, its frequency dependence decays as $1/f$ until it reaches the $1.6$~MHz roll-off of the low-pass filter applied during signal channelization. Conversely, in the case of phase noise, for frequency offsets above $10$~kHz it reaches a plateau at around $\approx200$~kHz, where the closed-loop gain of the Phase-Locked Loop (PLL) strongly attenuates the phase noise. This plateau is produced by the noise of the Voltage-Controlled Oscillator (VCO) present in the synthesizer's PLL\cite{HajiPLL}. Appendix~\ref{app:synthpn} provides a detailed analysis of the noise sources involved in the synthesis of the sampling clocks, to which the excess phase noise is attributed.
In contrast to the multiplicative sources, the additive noise spectral density which is expected to be dominated by the cold readout system was theoretically calculated using equation \ref{eq:addnsd}. As an example, for a typical cryogenic LNA with $T_{\mathrm{n}}=4$~K and assuming a probing power $P_{\mathrm{exc}}=-70$~dBm, the additive noise spectral density yields $S_{\sigma}(\Delta f) \approx -122.5$~dBc/Hz. The noise values shown in figure \ref{fig:readout-noise} will be used in the last section along with the \textmu MUX model (table \ref{tab:umuxparams}) to estimate the readout noise performance for this particular example. These noise values were measured at detector $f_{\mathrm{det}}\approx762$~Hz and modulation frequencies $f_{\mathrm{mod}}\approx62$~kHz and represented with black and green stars and magenta and cyan triangles in figure~\ref{fig:readout-noise}.

After all noise sources and metrics were introduced and an example was given, the error term in the determination of the transmission scattering parameter as a function of time $S_{21}(t)$ can be written as, 

\begin{equation}
     \delta S_{21}(t)=S_{21}(t)\gamma_{n}(t)e^{j\phi_n(t)}+\frac{\sigma_n(t)}{V_{\mathrm{exc}}}
     \label{eq:error}
\end{equation}

In the next section we will introduce the two most widely used demodulation domains and analyze the projection of the error term $\delta S_{21}(t)$ on them in order to evaluate the demodulation performance.

\subsection{\label{subsec:demod}Demodulation Domains and Schemes}

Slow variations in the flux, given either by a modulation $\varphi_{\mathrm{mod}}(t)$ or a detector $\varphi_{\mathrm{det}}(t)$, lead to variations of the resonator frequency response as described by equation \ref{eq:freqres} and depicted in figure \ref{fig:trajectories}. The most widely used domains in which these variations are demodulated are: 1) The resonator's phase measured from its rotating frame\cite{Gard2018,Becker_2019,Malnou} and 2) The amplitude of the transmission scattering parameter\cite{Karcher2022,Schuster2023}. Both are respectively defined as,

\begin{equation}
    \theta(f_{\mathrm{exc}},\varphi)=\arctan{\left\{\frac{\operatorname{Im}[S_{21}(f_{\mathrm{exc}},\varphi)]}{x_c-\operatorname{Re}[S_{21}(f_{\mathrm{exc}},\varphi)]}\right\}}
    \label{eq:resphase}
\end{equation}

\begin{equation}
    \gamma(f_{\mathrm{exc}},\varphi)=|S_{21}(f_{\mathrm{exc}},\varphi)|
    \label{eq:amp}
\end{equation}

The arbitrary selection of these two domains is based on the most commonly used demodulation types according to the literature and the availability of experimental results to validate the results of our simulations\cite{Malnou,Schuster2023}. A representation of both domains is shown at the left of figure \ref{fig:trajectories}. While both domains are used showing outstanding performance, there is not an available simulation framework that allows to compare their performance in the presence of readout system noise, especially multiplicative. Using equations \ref{eq:resphase} and \ref{eq:amp} we are able to define the projection of the error term $\delta S_{21}(t)$ in both domains. Thus, $\theta_n^{\sigma_n}(t)$, $\gamma_n^{\sigma_n}(t)$ correspond to resonator-phase and scattering-amplitude error due to additive noise $\sigma_n(t)$, while $\theta_n^{\phi_n}(t)$ and $\theta_n^{\gamma_n}(t)$ correspond to the resonator's phase error due phase $\phi_n(t)$ and amplitude $\gamma_n(t)$ noise respectively. Unlike the rest, amplitude noise $\gamma_n(t)$ is already projected into the scattering amplitude domain by definition. These projections are depicted in figure \ref{fig:trajectories} using shaded yellow, green, red and blue triangles. In the condition of small noise amplitudes, the geometrical projection of each noise source into the different domains can be calculated and the NSD derived yielding the analytical expressions shown in table \ref{tab:projectios}. A detailed analysis of these derivations are found in appendix \ref{app:projections}. The dependence of the NSD with $f_{\mathrm{exc}}$, $\varphi$ and $\Delta f$ was removed in order to avoid a complex notation.

\begin{table*}
\caption{\label{tab:projectios}. Spectral densities of additive and multiplicative noise sources projected in both demodulation domains. $S_{\theta}$ represents the flux spectral density in resonator's phase readout while $S_{\gamma}$ in scattering amplitude. The dependence of the parameters with $f_{\mathrm{exc}}$, $\varphi$ and $\Delta f$ was removed in order to avoid a complex notation.}
\begin{ruledtabular}
\begin{tabular}{cccc}
Readout Domain &Additive Noise $\sigma_n$ & Amplitude Noise $\gamma_n$& Phase Noise $\phi_n$\\ \hline
$S_{\theta}$&$\frac{1}{r^2}\frac{k_b T_{n}}{P_{\mathrm{exc}}}$&$\frac{\gamma^2\sin^2{(\phi+\theta)}}{r^2}S_{\gamma}$&$\frac{\gamma^2\cos^2{(\phi+\theta)}}{r^2} S_{\phi}$\\ \hline
$S_{\gamma}$&$\frac{k_b T_{n}}{P_{\mathrm{exc}}}$&$\gamma^2S_{\gamma}$&--\\
\end{tabular}
\end{ruledtabular}
\end{table*}

Expressions in table \ref{tab:projectios} corresponding to additive noise represent the easiest cases to analyze because they describe a constant radius noise cloud independent of the trajectory described by the resonator's response. Therefore, its projections either in resonator's phase or scattering amplitude are constant. Conversely, multiplicative noise projections depend on the position in the resonance circle $S_{21}(f_{\mathrm{exc}},\varphi)$ in which they are calculated (i.e. scattering amplitude $\gamma$, scattering phase $\phi$ and resonator's phase $\theta$). Particularly in these cases, there can be a condition for minimum or zero projection depending on the readout domain and readout parameters chosen. For example, in the case of readout amplitude noise $\gamma_n(t)$ there is no projection onto the resonator's phase domain when the readout phasor (red arrow in figure \ref{fig:trajectories}) is parallel to the circle radius ($\theta+\phi=0$). In contrast, for readout phase noise $\phi_n(t)$, there is no projection onto the resonator's phase domain when the readout phasor is tangent to the resonance circle ($\theta+\phi=\pi/2$). In addition, using the expressions in table \ref{tab:projectios} we conclude that additive noise is the dominant noise due to the fact that, when considering equal noise densities (i.e. $S_{\gamma}=S_{\phi}=\frac{k_B T_n}{P_{\mathrm{exc}}}$), multiplicative sources are scaled by factors lower than unity respect to the additive. Despite these expressions allowing us to calculate resonator's phase and scattering amplitude noise projections, an additional step is required in order to convert them to flux noise for a later comparison. This step depends on the demodulation scheme used and will be explained below.

\subsubsection{\label{subsubsec:opl}Open-Loop Demodulation}

When a detector signal is coupled to the SQUID through the input coil $L_{\mathrm{in}}$ (see figure \ref{fig:umux}), it creates a flux component $\varphi_{\mathrm{det}}(t)$ that is added to the modulation flux $\varphi_{\mathrm{mod}}(t)$. For a given constant $f_{\mathrm{exc}}$, $P_{\mathrm{exc}}$, and flux $\varphi_{\mathrm{mod}}$, small flux variations $\delta \varphi_{\mathrm{det}}(t)$ will lead to resonator's phase $\theta(t)$ and scattering amplitude $\gamma(t)$ variations described by,

\begin{equation}
    \theta(t) \approx \delta \varphi_{\mathrm{det}}(t) \frac{\partial \theta}{\partial \varphi}\bigg \rvert_{f_{\mathrm{exc}},P_{\mathrm{exc}},\varphi_{\mathrm{mod}}}
\end{equation}

\begin{equation}
    \gamma(t) \approx \delta \varphi_{\mathrm{det}}(t)\frac{\partial \gamma}{\partial \varphi}\bigg \rvert_{f_{\mathrm{exc}},P_{\mathrm{exc}},\varphi_{\mathrm{mod}}}
    \label{eq:gamma_op}
\end{equation}

\begin{figure}
\includegraphics[width=0.48\textwidth]{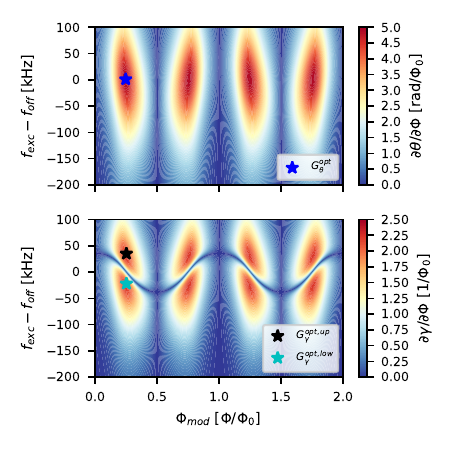}
\caption{\label{fig:op-gain} Open-loop gains corresponding to the \textmu MUX model presented in table \ref{tab:umuxparams} as a function of probe frequency $f_{exc}$ and modulation flux $\varphi_{mod}$ for a constant probing power $P_{\mathrm{exc}}=-70$~dBm. Top) Resonator's phase gain, blue star represents the maximum gain $G_\theta^{\mathrm{opt}}$. Bottom) Scattering amplitude gains, black and cyan stars represent the both upper $G_\gamma^{\mathrm{opt},\mathrm{up}}$ and lower $G_\gamma^{\mathrm{opt},\mathrm{low}}$ local maximum gains.}
\end{figure}

This readout scheme is called open-loop and the demodulated signals $\theta(t)$ and $\gamma(t)$ are scaled copies of $\delta \varphi_{det}(t)$. The scaling factors are denoted as open-loop gains, or transfer coefficients equivalently, being equal to the partial derivatives of the corresponding demodulation domain with respect to the normalized flux evaluated at $f_{\mathrm{exc}}$, $P_{\mathrm{exc}}$ and $\varphi_{\mathrm{mod}}$. Consequently, the maximum demodulated signal amplitudes are obtained when the maximum partial derivatives are achieved. Figure \ref{fig:op-gain} shows the simulated resonator's phase (top) and scattering amplitude (bottom) derivatives for the \textmu MUX model described in section \ref{tab:umuxparams} as a function of the modulation flux $\varphi_{\mathrm{mod}}$ and probing tone frequency $f_{\mathrm{exc}}$ for a specific probing power $P_{\mathrm{exc}}=-70$~dBm. The absolute maximum gain for the resonator's phase $G_\theta^{\mathrm{opt}}$ is represented by the blue star, while upper $G_\gamma^{\mathrm{opt},\mathrm{up}}$ and lower $G_\gamma^{\mathrm{opt},\mathrm{low}}$ local maximum open-loop gains with black and cyan stars respectively.

These gain factors allow us to calculate the resonator's phase and and scattering amplitude variations, but also can be used in the reverse way to derive the equivalent flux change for given resonator phase and amplitude variations. Therefore, the NSD of the noise projections listed in table \ref{tab:projectios} can be translated into an equivalent flux noise density for resonator's phase $\sqrt{S_{\Phi}^{\theta}}$ and scattering amplitude $\sqrt{S_{\Phi}^{\gamma}}$ readout using,

\begin{equation}
    \sqrt{S_{\Phi}^{\theta}} \approx \sqrt{S_{\theta}(\Phi)} \left| \frac{\partial \theta(\Phi)}{\partial \Phi}\right|^{-1}
    \label{eq:phase-opln}
\end{equation}

\begin{equation}
    \sqrt{S_{\Phi}^{\gamma}} \approx \sqrt{S_{\gamma}(\Phi)}\left| \frac{\partial \gamma(\Phi)}{\partial \Phi}\right|^{-1}
    \label{eq:gamma-opln}
\end{equation}

Simplifying the notation for reading purposes, the dependence of the noise projection with probing frequency and power was avoided only by preserving the explicit dependence with the magnetic flux $\Phi$. A similar expression for the flux noise $\sqrt{S_{\Phi}^{TLS}}$ can be found for TLS noise\cite{matesthesis,ahrensthesis}. Since it acts directly in the resonance frequency, it can be related to the flux using,

\begin{equation}
    \sqrt{S_{\Phi}^{TLS}} \approx \sqrt{S_{f_r}} \left| \frac{\partial f_r(\Phi)}{\partial \Phi}\right|^{-1}
    \label{eq:tls-opln}
\end{equation}

Where $\frac{\partial f_r(\Phi)}{\partial \Phi}$ is the derivative of equation \ref{eq:fr} with respect to the flux and $\sqrt{S_{f_r}}$ the frequency NSD. Although it is experimentally possible to determine the optimum flux bias necessary to obtain the maximum transfer coefficients, one of the main disadvantages of this scheme is that the information lies in the modulation side-bands located at a frequency offset equal to $f_{\mathrm{det}}$, where generally the phase $S_{\phi}(f_{det})$, amplitude $S_{\gamma}(f_{det})$ and TLS $S_{f_r}(f_{det})$ noise sources are considerably larger compared to the Johnson–Nyquist noise $\frac{k_B T_n}{P_{exc}}$, as it can be seen in figure \ref{fig:readout-noise} represented with green and black stars. Therefore, special care should be given to multiplicative and TLS noise sources in the case of bolometric applications where the detector signal is extremely slow and faint\cite{Geria2022,Dober2021}, or a different readout method to overcome this problem has to be used.

\subsubsection{\label{subsubsec:frm}Flux-Ramp Demodulation}

\begin{figure}
\includegraphics[width=0.48\textwidth]{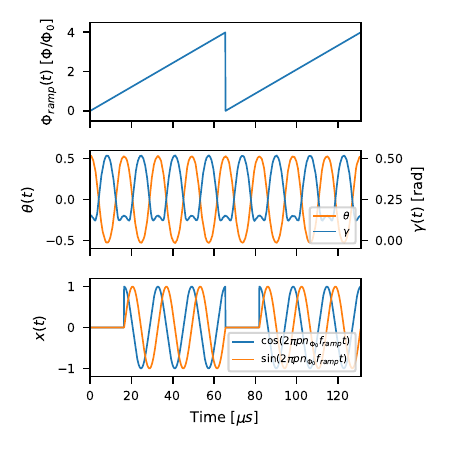}
\caption{\label{fig:s21demod} Signals associated with the flux-ramp demodulation process of the \textmu MUX model presented in table \ref{tab:umuxparams} for a particular $f_{\mathrm{exc}}$ and $P_{\mathrm{exc}}$. Top) Flux-Ramp Modulation signal spanning $n_{\Phi_0}=4$ with $f_{\mathrm{ramp}} \approx 15.25$~kHz. Middle) Resonator's phase and scattering amplitude signals before FRD. Bottom) Reference signals used for demodulation including the discarded period $n_{\mathrm{disc}}=1$ and using the first harmonic component $p=1$.}
\end{figure}

In addition to the aforementioned problem, random flux offsets trapped inside the SQUID loop would require individual flux tuning of every channel in order to achieve the maximum open-loop gain, re-introducing the multiplexing problem. A technique called Flux-Ramp Modulation\cite{Mates2012} (FRM) was introduced as a solution to overcome this limitation. In this technique, a common flux line is shared among all SQUIDS and a sawtooth-shaped flux with amplitude $M_{\mathrm{mod}}I_{\mathrm{mod}}=n_{\Phi_0} \Phi_0$ and reset frequency $f_{\mathrm{ramp}}$ is applied sweeping all the possible operating points (see upper plot in figure \ref{fig:s21demod}). Due to the SQUID periodic response, resonator's phase $\theta(t)$ and scattering amplitude $\gamma(t)$ will have the same periodicity equal to $f_{\mathrm{mod}}=n_{\Phi_0} f_{\mathrm{ramp}}$. Therefore, a considerably slow detector signal $\varphi_{det}(t)$ added to the the sawtooth-shaped modulation can be seen as a phase modulation, where $\varphi_{det}(t)$ determines the instantaneous phase changes of the periodic responses (middle plot in figure \ref{fig:s21demod}). As a consequence of FRM, the detector signal is up-converted to the frequency $f_{\mathrm{mod}}=n_{\Phi_0} f_{\mathrm{ramp}}$ (green region in figure \ref{fig:readout-noise}) avoiding in this way noise levels close to the probe tone frequency $f_{\mathrm{exc}}$. However, an additional phase demodulation step is required in order to recover the detector signal (Flux-Ramp Demodulation or FRD). In the context of this article we will analyze the flux demodulation noise performance using the linearity-improved quadrature demodulation\cite{SalumMuscheidFuster2023_1000162235}. The demodulated flux for the resonator's phase $\theta(t)$, and scattering amplitude $\gamma(t)$, demodulation can be written respectively as,

\begin{equation}
\varphi_{\theta}(t^{\prime})=\arctan\left[\frac{\int \sin(2 \pi p f_{\mathrm{mod}}t)w(t)\theta(t)dt}{\int \cos(2 \pi p f_{\mathrm{mod}}t)w(t)\theta(t)dt}\right]
\label{eq:phase-frd}
\end{equation}

\begin{equation}
\varphi_{\gamma}(t^{\prime})=\arctan\left[\frac{\int \sin(2 \pi p f_{\mathrm{mod}}t)w(t)\gamma(t)dt}{\int \cos(2 \pi p f_{\mathrm{mod}}t)w(t)\gamma(t) dt}\right]
\label{eq:gamma-frd}
\end{equation}

Here, $w(t)$ accounts for both the discarding window, where a number of SQUID periods $n_{\mathrm{disc}}$ is set to zero avoiding the sawtooth-shaped flux non-ideal transition and the window function used to attenuate the non-linearity components. An integer number $p$ is added allowing the demodulation of higher order FRM harmonics as we will use later in section \ref{sec:frm-optprams}. The bottom part of figure \ref{fig:s21demod} shows the cosine and sine reference signals used for FRD. Particularly in this example one of a total of four periods is discarded and a boxcar window is used. In this scheme, the integration is performed over a flux-ramp period setting an effective sampling rate of $f_{\mathrm{ramp}}$. Here, we use continuous time notation for explanatory purposes and $t'$ to explicitly show this implicit decimation. 

In contrast to open-loop demodulation, FRM sweeps all the possible flux values varying the open-loop gains from its maximum and minimum values. As a consequence, equations \ref{eq:gamma-opln} and \ref{eq:phase-opln} need to be integrated over a one flux-ramp period in order to get the demodulated noise level. As a natural solution, a gain-weighted average noise value may be used, but expressions \ref{eq:phase-opln}, \ref{eq:gamma-opln} and \ref{eq:tls-opln} diverge in the case of zero gain. Due to this, traditionally a Root-Mean-Square (RMS) gain over a flux-ramp period is defined\cite{Mates2012,Schuster2023} and flux noise density in the case of resonator's phase and scattering amplitude is calculated using,

\begin{equation}
    \sqrt{S_{\Phi}^{\theta}} \approx \sqrt{S_{\theta}} \left\{\frac{1}{n_{\Phi_0} \Phi_0} \int_{n_{\mathrm{disc}}\Phi_0}^{n_{\Phi_0}\Phi_0} \left( \frac{\partial \theta(\Phi)}{\partial \Phi}\right)^{2} d\Phi \right\}^{-\frac{1}{2}}
    \label{eq:phase-frmnoise}
\end{equation}

\begin{equation}
    \sqrt{S_{\Phi}^{\gamma}} \approx \sqrt{S_{\gamma}} \left\{\frac{1}{n_{\Phi_0}\Phi_0} \int_{n_{\mathrm{disc}}\Phi_0}^{n_{\Phi_0}\Phi_0} \left( \frac{\partial \gamma(\Phi)}{\partial \Phi}\right)^{2} d\Phi \right\}^{-\frac{1}{2}}
    \label{eq:gamma-frmnoise}
\end{equation}

Similarly for TLS, a flux noise density can be calculated as well,

\begin{equation}
    \sqrt{S_{\Phi}^{TLS}} \approx \sqrt{S_{f_r}} \left\{ \frac{1}{n_{\Phi_0}\Phi_0} \int_{n_{\mathrm{disc}}\Phi_0}^{n_{\Phi_0}\Phi_0} \left( \frac{\partial f_r(\Phi)}{\partial \Phi}\right)^{2} d\Phi \right\}^{-\frac{1}{2}}
\end{equation}

Where expressions between brackets are the gain-mean squares over a flux-ramp period and $S_{\theta}$, $S_{\gamma}$ and $S_{TLS}$ are evaluated at $f_{\mathrm{mod}}$. Due to the fact that we are discarding $n_{\mathrm{disc}}$ periods from a total of $n_{\Phi_0}$ and applying a window function, we expect a degradation equal to $\sqrt{\kappa/\alpha}$, where $\alpha=(n_{\Phi_0}-n_{\mathrm{disc}})/n_{\Phi_0}$. A $\kappa$ factor was added to account for the degradation caused by the window applied. When boxcar window is used $\kappa=1$. While these expressions give a rough estimation of the noise level, they have several limitations that this work aims to overcome. Unlike additive and TLS noise, these expressions do not apply to the case of multiplicative noise where both spectral densities and gains are flux dependent (i.e. $\sqrt{S_{\theta,\gamma}(\Phi)}$ ). Another limitation is that, due to the non-sinusoidal resonator's phase and scattering amplitude responses, the information is not only contained in a single side-band, but spread out in harmonics of $f_{\mathrm{mod}}$ which cannot be demodulated using the quadrature demodulation method. Additionally, an extra challenge lies in the analytical calculation of the derivatives with respect to the flux, for the given conditions of readout power and frequency using the model presented in section \ref{sec:umuxmodel} and their integration over a flux-ramp period. This is remarkable especially when multidimensional optimization criteria must be found. That is why in this work the evaluation of the optimal readout parameters that match the lowest noise condition in both domains and demodulation methods will be performed by means of numerical simulations.

\section{\label{sec:sysnoise}System Noise Simulation}

In this section we present several simulations performed in order to evaluate the demodulated noise for different readout conditions. Previously, for sake of simplicity a continuous-time signal model of the readout system was introduced. From now on, we will migrate to its discrete-time equivalent using adequate sampling frequency $f_s$ that satisfies the sampling theorem. This migration is not only motivated by the numerical simulation, but also because the demodulation is typically done in the digital domain using Software-Defined Radio (SDR) Systems\cite{Yu2023,Gard2018,Ferreyro2023,Karcher2022}. The readout system simulation framework was implemented completely in Python and making use of several available libraries. Table \ref{tab:readoutparams} enumerates the parameters used in the following simulations.

The general simulation procedure to obtain a demodulated signal time-trace can be summarized as follows:

\begin{enumerate}
    \item The total modulation flux $\varphi_{\mathrm{tot}}[m]$ is created adding the modulation flux $\varphi_{\mathrm{mod}}[m]$ to the detector flux $\varphi_{\mathrm{det}}[m]$.
    \item Given fixed $f_{\mathrm{exc}}$ and $P_{\mathrm{exc}}$, $f_{r}[m]$ is iteratively calculated using equation \ref{eq:fr} and $f_{TLS}[m]$ is added.
    \item $S_{21}[m]$ is calculated using equation \ref{eq:freqres} for a particular $f_{\mathrm{exc}}$, low-pass filtered and the readout noise sources $\sigma_n[m]$, $\phi_n[m]$ and $\gamma_n[m]$ included.
    \item Finally, and after a new step of signal conditioning, $\theta[m]$ and $\gamma[m]$ are available to be demodulated using some of the methods presented in previous section.
\end{enumerate}

\begin{table}
\caption{\label{tab:readoutparams}Readout system parameters used during demodulated noise simulations.}
\begin{ruledtabular}
\begin{tabular}{lcr}
Parameter&Value&Unit\\
\hline
$f_{s}$& $\approx$7.82&  MHz\\
$N_{s}$& 2\textsuperscript{20} &  Samples\\
$BW_{\mathrm{readout}}$& 2 & MHz\\
$f_{\mathrm{det}}$ & $\approx$762.90 & Hz\\
$A_{\mathrm{det}}$ & 1 & m$\Phi_0$\\
$f_{\mathrm{ramp}}$& $\approx$15.25 &  kHz\\
$n_{\Phi_0}$& 4 &  -\\
$n_{\mathrm{disc}}$ & 1 & -\\
$p$ & 1 & -\\
$w[n]$ & Hamming& -\\
\end{tabular}
\end{ruledtabular}
\end{table}

Throughout the simulations presented in this section some simplifications were adopted. SDR systems exhibit additive and multiplicative noise levels dependent on the attenuation or gain required to obtain the desired probing tone power levels $P_{\mathrm{exc}}$ at the \textmu MUX input or the appropriate signal levels at the receiver input. In order to simplify the analysis, we will assume that both the additive and multiplicative noise levels shown in Figure~\ref{fig:readout-noise} (i.e. $T_{n}$, $S_{\phi}$ and $S_{\gamma}$) are constant over the power range in which the simulations were performed. Although these simplifications are not applicable to all SDR systems, they are applicable to the case of the SDR system characterized in the appendix~\ref{app:readnoisecharact}.

\subsection{\label{sec:opl-noise}Open-Loop Demodulation Noise Performance}

Using our simulation framework, we replicated the measurement process that is carried out in the laboratory during characterization. This allowed us to not only evaluate the noise performance, but also to know the required functionalities of our SDR readout system. For the determination of the noise density in the open-loop scheme described by equations \ref{eq:phase-opln} and \ref{eq:gamma-opln}, we must know the value of the open-loop gains (transfer coefficients) which depend on the readout frequency $f_{\mathrm{exc}}$ and power $P_{\mathrm{exc}}$ as well as the applied constant modulation flux $\varphi_{\mathrm{mod}}$. First, for each value of frequency and power, one period of the transmission scattering parameter $S_{21}(f_{\mathrm{exc}},P_{\mathrm{exc}},\varphi_{\mathrm{mod}})$ was calculated sweeping $\varphi_{\mathrm{mod}}$. Then, we get resonator's phase $\theta(f_{\mathrm{exc}},P_{\mathrm{exc}},\varphi_{\mathrm{mod}})$ and scattering amplitude $\gamma(f_{\mathrm{exc}},P_{\mathrm{exc}},\varphi_{\mathrm{mod}})$ to consequently obtain the partial derivatives with respect to flux as can be seen in figure \ref{fig:op-gain}. For the next process, we stored the maximum gains $G_{\theta}^{\mathrm{opt}}(f_{\mathrm{exc}},P_{\mathrm{exc}})$ and $G_{\gamma}^{\mathrm{opt}}(f_{\mathrm{exc}},P_{\mathrm{exc}})$ along with the corresponding fluxes $\varphi_{\mathrm{mod}}^{\mathrm{opt}}$ and resonance circle parameters (i.e. radius $r$ and center $x_c$). Resonance circle parameters determine the rotating frame required to calculate the resonator's phase using equation $\ref{eq:resphase}$ when detector signal is present. This calibration procedure allows us to get all the parameters required to obtain a demodulated signal time-trace.

\subsubsection{\label{sec:opl-optdparams}Optimum Readout Parameters for Open-Loop}

\begin{figure}
\includegraphics[width=0.48\textwidth]{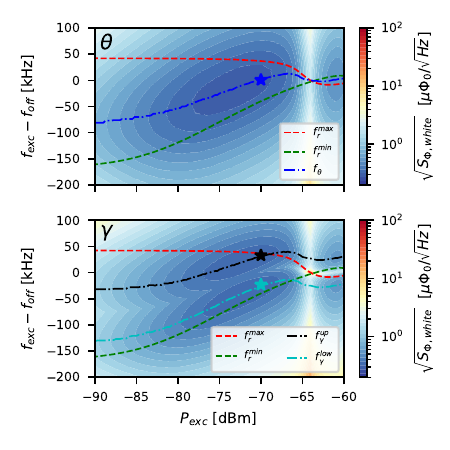}
\caption{\label{fig:opl-noise} Open-loop demodulated white noise flux density $\sqrt{S_{\Phi,white}}$ for both demodulation domains as a function of probe tone frequency $f_{\mathrm{exc}}$ and power $P_{\mathrm{exc}}$ for an additive noise temperature of $T_n=4$~K. Red and green dashed lines represent maximum $f_{r}^{\mathrm{max}}$ and minimum $f_{r}^{\mathrm{min}}$ resonance frequencies respectively. Top) Resonator's phase demodulation, dashdotted blue $f_{\theta}$ line is the optimum trajectory to achieve minimum noise. Bottom) Scattering amplitude demodulation, dashdotted black and cyan curves are the optimum trajectories to achieve both upper $f_{\gamma}^{\mathrm{up}}$ and lower $f_{\gamma}^{\mathrm{low}}$ local minimum noise values.}
\end{figure}

In order to find the optimum readout parameters, for each frequency $f_{\mathrm{exc}}$ and power $P_{\mathrm{exc}}$, a small detector signal $\varphi_{\mathrm{det}}[m]$ was added to the fluxes corresponding to the maximum gains $\varphi_{\mathrm{mod}}^{\mathrm{opt}}$. The transmission parameter calculated and the readout noise sources are included. Consequently, depending on the selected domain, resonator's phase $\theta[m]$ and scattering amplitude $\gamma[m]$ time-traces calculated and $\varphi_{\theta}[m]$ and $\varphi_{\gamma}[m]$ derived dividing them by their correspondent optimum open-loop gain coefficients. By means of the Welch’s method\cite{welch}, the white flux noise spectral density $\sqrt{S_{\Phi,white}}$ was calculated. The results of these simulations for both domains are shown in figure \ref{fig:opl-noise} for the case of $T_n=4$~K, typical noise equivalent temperature of cryogenic HEMT amplifier\cite{Rao2020}. Red and green dashed lines in figure \ref{fig:opl-noise} represent maximum and minimum resonance frequencies according to equation \ref{eq:fr}, while dashdotted blue, black and cyan represent the minimum noise $(P_{\mathrm{exc}},f_{\mathrm{exc}})$ trajectories. These three trajectories describe the position followed by the optimum open-loop gains previously presented in figure \ref{fig:op-gain}. Colored stars are the local noise minimums at $P_{\mathrm{exc}}=-70$~dBm, exactly at the same positions where the maximum gains were calculated in figure \ref{fig:op-gain}. As we previously mentioned, only additive noise was considered. Therefore, the optimum readout parameters are valid for this condition.

\begin{figure}
\includegraphics[width=0.48\textwidth]{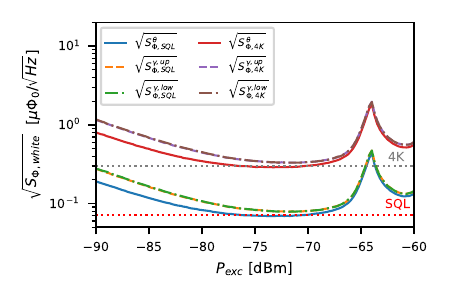}
\caption{\label{fig:opl-opt} Open-loop demodulated minimum white noise flux density $\sqrt{S_{\Phi,white}}$ for every demodulation domain as a function of power $P_{\mathrm{exc}}$ for both SQL and $T_n=4$~K cases. Solid lines correspond to minimum noise for resonator's phase readout while dashed and dashdotted lines are upper and lower local minimums for scattering amplitude readout. Dotted grey and red lines are the minimum noise values for resonator's phase demodulation in the Standard Quantum Limit (SQL) and $T_n=4$~K conditions evaluated at $P_{\mathrm{exc}}=-70$~dBm.}
\end{figure}

As can be seen in figure \ref{fig:opl-noise}, the noise minimums for low powers always lie within the maximum and minimum resonant frequencies and the optimum conditions for scattering amplitude are symmetrically located on both sides of the optimum condition for the resonator's phase. In order to compare the demodulated noise values for both demodulation domains, the minimum noise levels corresponding to these trajectories were simulated again for two different additive noise temperatures. In line with new amplification technologies, we performed simulations in the Standard Quantum Limit\cite{sql,Malnou} (SQL) condition with $T_n=h f_{\mathrm{exc}}/k_B$ along with the $T_n=4$~K previously used. The results are plotted together in figure \ref{fig:opl-opt}. For these cases, flux noise levels can be estimated using the additive noise version of equations \ref{eq:phase-opln} and \ref{eq:gamma-opln},

\begin{equation}
    \sqrt{S_{\Phi}^{\theta}} \approx \frac{1}{r} \sqrt{\frac{k_B T_n}{P_{\mathrm{exc}}}} \left| \frac{\partial \theta}{\partial \Phi}\right|^{-1}
    \label{eq:phase_tn}
\end{equation}

\begin{equation}
    \sqrt{S_{\Phi}^{\gamma}} \approx \sqrt{\frac{k_B T_n}{P_{\mathrm{exc}}}}\left| \frac{\partial \gamma}{\partial \Phi}\right|^{-1}
    \label{eq:amp_tn}
\end{equation}

In accordance to these equations, at low power the transfer coefficients remain constant and the noise level decreases proportionally to $\sqrt{S_{\Phi,white}} \propto 1/\sqrt{P_{\mathrm{exc}}}$ until it reaches a plateau in which both open-loop gain values reach a level of around $P_{\mathrm{exc}}\approx -63$~dBm  where the \textmu MUX becomes insensitive to flux variations. Using $P_{\mathrm{exc}}=-70$~dBm, simulated optimum gain values yielded $G_{\theta}^{\mathrm{opt}}=4.90$ $rad/\Phi_0$ and $G_{\gamma}^{\mathrm{opt,up}}=G_{\gamma}^{\mathrm{opt,low}}=2.21/\Phi_0$ (see figure \ref{fig:op-gain}). Thus, for a resonance circle radius $r \approx Q_l/2Q_c\approx 0.49$ the calculated noise values results are $\sqrt{S_{\Phi,SQL}^{\theta}} \approx 0.071 \mu\Phi_0/\sqrt{Hz}$ and $\sqrt{S_{\Phi,SQL}^{\gamma}} \approx 0.08 \mu\Phi_0/\sqrt{Hz}$ for the SQL, in contrast to $\sqrt{S_{\Phi,4K}^{\theta}} \approx 0.30 \mu\Phi_0/\sqrt{Hz}$ and $\sqrt{S_{\Phi,4K}^{\gamma}} \approx 0.34 \mu\Phi_0/\sqrt{Hz}$ with $T_n=4$~K, in well agreement with the simulations. Dotted red and grey lines represent the aforementioned noise values in the case of resonator's phase readout. Leaving aside technical difficulties that may be encountered when implementing the demodulation of one of the domains, which will be discussed in the next subsections, there is no strong argument based on noise performance to decide for either domain. This is given by the non-appreciable noise differences in the simulation results. Therefore, we will evaluate the impact of the remaining noise sources in order to find considerable differences that would indicate the existence of a preferential domain.

\subsubsection{\label{sec:opl-multip}TLS and Multiplicative Noise Impact for Open-Loop}

\begin{figure}
\includegraphics[width=0.48\textwidth]{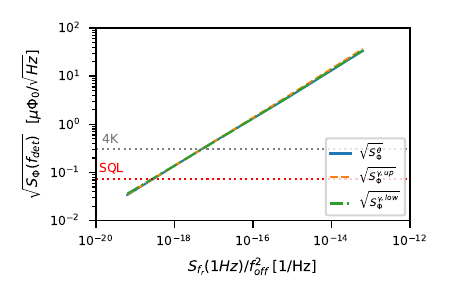}
\caption{\label{fig:opl-equivtls} Open-loop demodulated white noise flux density $\sqrt{S_{\Phi,white}}$ as a function of the Two-Level Systems (TLS) noise level of $S_{f_r}(1 Hz)/f_{\mathrm{off}}^2$ for both demodulation domains in the minimum additive noise condition for a constant $P_{\mathrm{exc}}=-70$~dBm. As a reference, dotted red and grey lines represent the minimum noise values for resonator's phase demodulation in the case of the Standard Quantum Limit (SQL) and $T_n=4$~K respectively.}
\end{figure}

In order to quantify the impact of the non-additive noise sources, we performed individual simulations for each noise source under the optimum readout parameters for additive noise, represented with stars in figure \ref{fig:opl-noise}. As mentioned earlier, Two-Level Systems noise is one of the most common noise sources and particularly important in open-loop readout due to its frequency dependence\cite{Gao2007}. Here, TLS noise was represented by its fractional frequency density $S_{f_r}/f_{\mathrm{off}}^2$ with a frequency slope of $1/f$ and taking values from $6.25\cdot 10^{-20}$~Hz$^{-1}$ to $6.25\cdot 10^{-14}$~Hz$^{-1}$ when evaluated at $1$~Hz. Due to the TLS noise saturation depending on the readout power\cite{Gao2007}, it is worth to mention that TLS noise is specified for $P_{\mathrm{ext}}=-70$~dBm. Simulations results for each domain measured at detector frequencies of $f_{\mathrm{det}}\approx762$~Hz are shown in figure \ref{fig:opl-equivtls}. As in the previous plots, the minimum noise values for the SQL and $T_n=4$~K were included as reference.

Results in figure \ref{fig:opl-equivtls} allows us to set maximum TLS noise levels of $S_{f_r}(1 Hz)/f_{\mathrm{off}}^2=3\cdot 10^{-18}$~Hz$^{-1}$ and $S_{f_r}(1 Hz)/f_{\mathrm{off}}^2=6.6\cdot 10^{-17}$~Hz$^{-1}$ for the SQL and $T_n=4$~K respectively avoiding the TLS as dominant noise source, but once again, showing no preferential demodulation domain.

\begin{figure}
\includegraphics[width=0.48\textwidth]{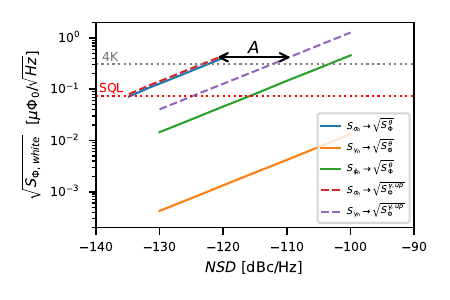}
\caption{\label{fig:opl-equiv} Open-loop demodulated white flux noise density $\sqrt{S_{\Phi,white}}$ for every demodulation domain as a function of the noise spectral densities (NSD) for each noise source in the minimum additive noise condition for a constant $P_{\mathrm{exc}}=-70$~dBm. Dashed lines correspond to scattering amplitude domain, while solid lines are for resonator's phase readout. Grey and Red dotted lines are plotted as a reference and are equal to the white noise values for the Standard Quantum Limit (SQL) and for $T_\mathrm{n}=4$~K for resonator's phase readout. As an example, the black arrow represents the amplitude noise rejection factor $A$ in scattering amplitude domain readout.}
\end{figure}

Contrary to the TLS noise, multiplicative and additive noise sources can be compared directly in the same NSD units. Therefore, the simulations were performed as a function of the NSD values for each noise source in the same conditions as before using $P_{\mathrm{exc}}=-70$~dBm. Equivalent noise temperatures starting from the SQL with $T_n=h f_{\mathrm{exc}}/k_B \approx 0.24$~K, to $T_n=8$~K, corresponding to NSDs from $-135$~dBc/Hz to $-120$~dBc/Hz were used in the case of additive noise. While for phase and amplitude readout noise, NSD going from $-130$~dBc/Hz to $-100$~dBc/Hz. Since both local minimums for scattering amplitude yield the same demodulated noise for open-loop demodulation, only the upper local minimum is considered in the following analysis. The results of these simulations as a function of the NSD are shown in figure \ref{fig:opl-equiv}. As demonstrated previously, resonator's phase readout lead to a slightly lower noise than scattering amplitude readout for the same NSD. In addition, as expected from table \ref{tab:projectios}, noise levels scales proportional to NSD with $\sqrt{S_{\Phi}} \propto \sqrt{NSD}$, being additive noise the dominant noise source compared with multiplicative sources for equal NSD values. When considering non-correlated noise sources acting together, the total system noise is the quadrature summation all individual contributions. Therefore, we defined a rejection factor $A$ for each noise source and demodulation domain equal to the NSD difference with respect to the additive NSD that would result in the same demodulated noise level. This factor is represented with a black arrow in figure \ref{fig:opl-equiv} for the case of scattering amplitude demodulation and is equal to the distance in NSD units between parallel dashed lines for a constant flux noise density. Thus, for scattering amplitude demodulation there is a $\approx10$~dB rejection of amplitude noise. In contrast, using resonator's phase demodulation, there are $\approx19$~dB and more than $40$~dB rejection values for phase and amplitude noise respectively. Particularly, scattering amplitude demodulation is more sensitive than resonator's phase demodulation to amplitude noise because the readout tone is positioned further away from the resonance where its projection is considerably greater compared to phase noise and the scattering amplitude open-loop gain lower. On the contrary, resonator's phase demodulation is more sensitive to phase noise because the maximum open-loop gain requires the probing tone in resonance where phase noise projection is maximum (see figure \ref{fig:trajectories}).

\subsubsection{\label{sec:sdr-example-opl}Optimum Domain for Open-loop}

\begin{figure}
\includegraphics[width=0.48\textwidth]{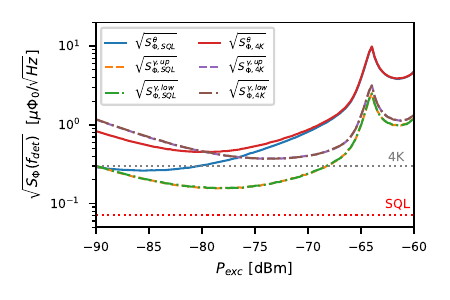}
\caption{\label{fig:readout-example-opl} Open-loop demodulated flux noise density $\sqrt{S_{\Phi}(f_{\mathrm{det}})}$ for every demodulation domain as a function of power $P_{\mathrm{exc}}$ for both SQL and $T_n=4$~K cases. Solid lines correspond to minimum noise for resonator's phase readout, while dashed and dashdotted lines are upper and lower local minimums for scattering amplitude readout. Dotted grey and red lines are the minimum noise values for resonator's phase demodulation in the Standard Quantum Limit (SQL) and $T_n=4$~K conditions evaluated at $P_{\mathrm{exc}}=-70$~dBm.}
\end{figure}

Previous results suggest that maximum sensitivity is achieved in the resonator's phase domain due to the lower demodulated noise level in the additive noise case and high rejection to multiplicative noise sources compared to scattering amplitude domain. Unfortunately, this is not a realistic scenario in the case of open-loop readout. Phase and amplitude noise spectral components (equation \ref{eq:powerlaw}) usually grow rapidly for small frequency offsets far exceeding additive noise, impacting directly in the detector signal. As an example, in order to determine the expected open-loop performance of a real SDR readout system\cite{zcu216}, the measured noise values shown in figure \ref{fig:readout-noise} were included in our simulation framework. Since the TLS noise did not show a preferential readout region and we assume a proper resonator design, it was not included in this case. The results of the demodulated noise for every domain for both SQL and $T_n=4$~K additive noise conditions are shown in figure \ref{fig:readout-example-opl}.

Clearly, the demodulated noise is dominated by the multiplicative noise and its effect is most noticeable at high powers and near the SQL\cite{sql}. In the case of the resonator's phase domain, it's noise is completely dominated by the multiplicative noise values of $S_{\phi}(f_{\mathrm{det}})\approx-96$~dBc/Hz and $S_{\gamma}(f_{\mathrm{det}})\approx-116$~dBc/Hz for phase and amplitude respectively. Both are considerably larger than the additive noise at the detector frequency  $S_{\sigma}(f_{\mathrm{det}})$ (see colored stars in figure \ref{fig:readout-noise}). Noise levels at $P_{\mathrm{exc}}=-70$~dBm are $\sqrt{S_{\Phi}^{\theta}(f_{\mathrm{det}})} \approx 0.82 \mu\Phi_0/\sqrt{Hz}$ and $\sqrt{S_{\Phi}^{\gamma}(f_{\mathrm{det}})} \approx 0.24 \mu\Phi_0/\sqrt{Hz}$ for the SQL, while $\sqrt{S_{\Phi}^{\theta}(f_{\mathrm{det}})} \approx 0.87 \mu\Phi_0/\sqrt{Hz}$ and $\sqrt{S_{\Phi}^{\gamma}(f_{\mathrm{det}})} \approx 0.40 \mu\Phi_0/\sqrt{Hz}$ for the $T_\mathrm{n}=4$~K case. All of them are consistent with the values shown in figure \ref{fig:opl-equiv} for their respective NSD levels.

While phase noise can be reduced at lower frequencies using an ultra-low phase noise synthesizer as suggested in appendix \ref{app:synthpn}, and TLS can be strongly reduced properly choosing resonator materials and geometry\cite{Gao2007}, they are still a problem specially in the case of bolometric applications. Therefore, the FRM technique mentioned above in \ref{subsubsec:frm}, can be used to mitigate these effects along with other limitations previously mentioned.

\subsection{\label{sec:frm-noise}Flux-Ramp Demodulation Noise Performance}

Similarly to the open-loop demodulation, we replicated the noise characterization procedure of a real \textmu MUX device. For the determination of the noise density in the FRD scheme described by equations \ref{eq:gamma-frd} and \ref{eq:phase-frd}. First, we generated the modulation flux $\varphi_{\mathrm{mod}}[m]$ and the detector signal $\varphi_{\mathrm{det}}[m]$ using the parameters described in table \ref{tab:readoutparams}. Both fluxes were added, $f_r[m]$ and the transmission parameter $S_{21}(f_{\mathrm{exc}},P_{\mathrm{exc}},\varphi_{\mathrm{mod}})[m]$ consequently iteratively calculated. Later, using the circle parameters previously stored, resonator's phase $\theta(f_{\mathrm{exc}},P_{\mathrm{exc}},\varphi)[m]$ and scattering amplitude $\gamma(f_{\mathrm{exc}},P_{\mathrm{exc}},\varphi_{\mathrm{mod}})[m]$ time traces were obtained. Finally, we apply the FRD for both domains in order to get the detector signal time-traces $\varphi_{\theta}[n]$ and $\varphi_{\gamma}[n]$. Here, we used $n$ instead of $m$ as a consequence of the decimation produced by the FRD\cite{Mates2012,SalumMuscheidFuster2023_1000162235} (see figure \ref{fig:s21demod}).

\subsubsection{\label{sec:frm-optprams}Optimum Readout Parameters for FRD}

\begin{figure}
\includegraphics[width=0.48\textwidth]{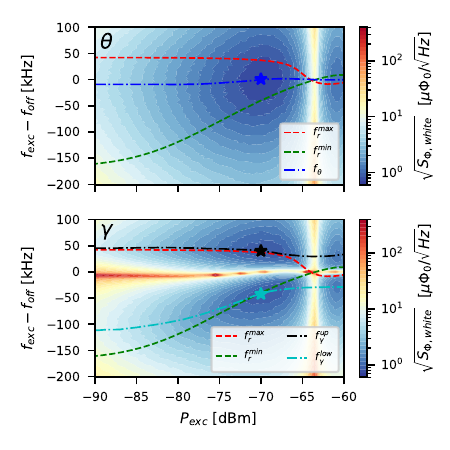}
\caption{\label{fig:frm-noise} Flux-ramp demodulated white flux noise density $\sqrt{S_{\Phi,white}}$ for both demodulation domains as a function of probe tone frequency $f_{\mathrm{exc}}$ and power $P_{\mathrm{exc}}$. Top) Resonator's phase demodulation, red and green dashed lines represent maximum $f_{r}^{\mathrm{max}}$ and minimum $f_{r}^{\mathrm{min}}$ resonance frequencies respectively. Dashdotted blue lines $f_{\theta}$ show the optimum trajectory to achieve minimum noise. Bottom) Scattering amplitude demodulation, dashdotted black and cyan curves are the optimum trajectories to achieve both upper $f_{\gamma}^{\mathrm{up}}$ and lower $f_{\gamma}^{\mathrm{low}}$ local minimum noise values. Hamming windows function with $n_{\mathrm{disc}}=1$ and $n_{\Phi_0}=4$ were used.}
\end{figure}

The demodulated detector time-traces $\varphi_{\mathrm{det}}[n]$ for both domains were calculated as a function of the frequency $f_{\mathrm{exc}}$ and power $P_{\mathrm{exc}}$ considering only additive noise with $T_\mathrm{n}=4$~K. By means of the Welch’s method\cite{welch}, the white flux noise spectral density $\sqrt{S_{\Phi,white}}$ was calculated. The results of these simulations for both domains are shown in figure \ref{fig:frm-noise}. Red and green dashed lines represent maximum and minimum resonance frequencies, while dashdotted blue, black and cyan the minimum noise $(P_{\mathrm{exc}},f_{\mathrm{exc}})$ trajectories. Colored stars are the absolute noise minimums at $P_{\mathrm{exc}} \approx -70$~dBm for both domains.

In principle, we would assume that the noise minima will also follow the trajectories previously found for the open-loop scenario (figure \ref{fig:opl-noise}). But this is not the case, the trajectories mainly differ, only approaching each other at high powers. This behaviour is mainly determined by the spectral components of the resonator's phase and scattering amplitude time-traces described by their Fourier series as,

\begin{equation}
    \theta(t) \approx \Theta_0 + \sum_{p=1}^{\infty} \Theta_p \sin(2 \pi p f_{\mathrm{mod}}t + \lambda_p)
\end{equation}

\begin{equation}
    \gamma(t) \approx \Gamma_0 + \sum_{p=1}^{\infty} \Gamma_p \sin(2 \pi p f_{\mathrm{mod}}t + \epsilon_p)
\end{equation}

\begin{figure}
\includegraphics[width=0.48\textwidth]{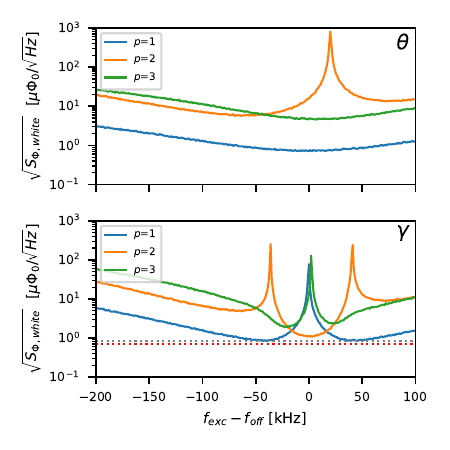}
\caption{\label{fig:harmonics} Flux-ramp demodulated white flux noise density $\sqrt{S_{\Phi,white}}$ as a function of $f_{\mathrm{exc}}$ for every demodulated domain and selected FRM component $f_{\mathrm{dem}}=p \cdot f_{\mathrm{mod}}$ for constant $P_{\mathrm{exc}}=-70$~dBm and additive noise temperature of $T_n=4$~K. Top) Resonator's phase demodulation. Bottom) Scattering amplitude demodulation. Red and grey lines show resonator’s phase and scattering amplitude noise minimums respectively. Hamming window functions with $n_{\mathrm{disc}}=1$ and $n_{\Phi_0}=4$ were used.}
\end{figure}

Where the fundamental frequency is $f_{\mathrm{mod}}=n_{\Phi_0} f_{\mathrm{ramp}}$ (also called SQUID frequency) and $\Theta_p$, $\Gamma_p$, $\lambda_p$, $\epsilon_p$ amplitude and phase spectral coefficients. Based on the FRM\cite{Mates2012}, we can assume that the detector signal changes the instantaneous frequency of each harmonic component $f_{\mathrm{dem}}=p \cdot f_{\mathrm{mod}}$ and consequently each component is phase modulated. However, while the modulation index is given by the detector signal, the carrier power depends on the amplitude spectral components of each domain (i.e. $\Theta_p$ and $\Gamma_p$). Due to the fact that the first component is usually demodulated, noise values increase when either $\Theta_p$ or $\Gamma_p$ decreases. In order to probe this, we swept $f_{\mathrm{exc}}$ at constant power of $P_{\mathrm{exc}}=-70$~dBm, and the detector signal was demodulated for each harmonic component changing $p$ in equations $\ref{eq:phase-frd}$ and $\ref{eq:gamma-frd}$. Only one period was discarded considering that the flux-ramp transient lasts less than one period of the highest modulating frequency harmonic. The results of the simulation are shown in figure \ref{fig:harmonics}.

Blue traces in both panels of figure \ref{fig:harmonics} correspond to the demodulation of the first harmonic component $p=1$, consistent with the results of figure \ref{fig:frm-noise}. Dotted red and grey lines are the minimum demodulated white flux noise for resonator's phase and scattering amplitude respectively, achieved at conditions represented by the blue, black an cyan stars. Like in the open-loop scheme, resonator's phase domain gives the lowest possible noise although the difference with scattering amplitude demodulation is not considerable. In the case of the resonator’s phase, we can see that demodulated noise for the first component is always lower than the rest. On the contrary, in the case of scattering amplitude readout, when the probe tone $f_{\mathrm{exc}}$ is close to the unaltered resonance frequency $f_{\mathrm{off}}$ noise diverges for the first and third demodulated component while for the second there is a minimum in accordance to results presented by Schuster\cite{Schuster2023}. Therefore, the absolute minimum noise conditions imply the demodulation of the first harmonic at its maximum amplitude (i.e. $\Theta_1^{max}$ and $\Gamma_1^{max}$). Since the total power is spread out between harmonics, this is fulfilled when the sum of the powers of the other harmonics reach their minimums. On the other hand, for a fixed $f_{\mathrm{exc}}$, the spectral components are determined by $P_{\mathrm{exc}}$\cite{Wegner_2022}. When power increases $P_{\mathrm{exc}}$, both $\beta_{\mathrm{eff}}$ and $\eta$ decrease, leading again to a sinusoidal response in resonator's phase and scattering amplitude, until there is no response (i.e. $\Theta_p \approx \Gamma_p \approx 0$). Taking advantage of the fact that resonator's phase and scattering amplitude behave as sinusoidal signals at high powers, equations \ref{eq:phase-frmnoise} and \ref{eq:gamma-frmnoise} in the additive noise case can be easily calculated considering $\theta(\Phi)\approx \Theta_1 \sin(2 \pi \Phi/\Phi_0)$ $\gamma(\Phi)\approx \Gamma_1 \sin(2 \pi \Phi/\Phi_0)$ yielding to\cite{Malnou},

\begin{equation}
    \sqrt{S_{\Phi}^{\theta}} \approx \frac{1}{r} \frac{\sqrt{2\kappa/\alpha}}{2 \pi \Theta_1} \sqrt{\frac{k_B T_n}{P_{\mathrm{exc}}}}
    \label{eq:phase_tn_frm}
\end{equation}

\begin{equation}
    \sqrt{S_{\Phi}^{\gamma}} \approx \frac{\sqrt{2\kappa/\alpha}}{2 \pi \Gamma_1} \sqrt{\frac{k_B T_n}{P_{\mathrm{exc}}}}
    \label{eq:gamma_tn_frm}
\end{equation}

\begin{figure}
\includegraphics[width=0.48\textwidth]{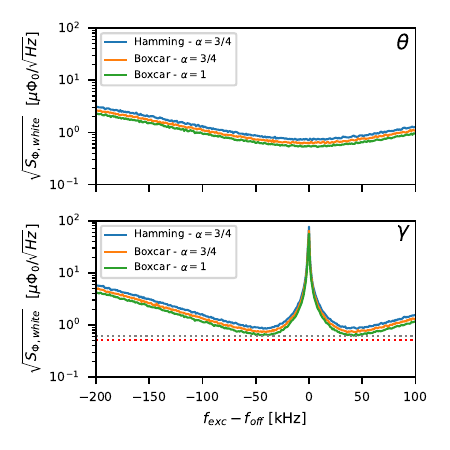}
\caption{\label{fig:windows} Flux-ramp demodulated white flux noise density $\sqrt{S_{\Phi,white}}$ as a function of $f_{\mathrm{exc}}$ for different window functions $w(t)$ and number of discarded periods $n_{\mathrm{disc}}$ with constant $P_{\mathrm{exc}}=-70$~dBm and additive noise temperature of $T_\mathrm{n}=4$~K. Top) Resonator's phase and Bottom) Scattering amplitude. Red and grey lines show phase and amplitude noise minimums respectively for a Boxcar window function, without periods discarded.}
\end{figure}

To verify the validity of these expressions, to later use them to estimate system temperature $T_{\mathrm{sys}}$, first we numerically determined the amplitude of the fundamental resonator's phase and scattering amplitude components yielding $\Theta_1 \approx 0.63$~rad and $\Gamma_1^{\mathrm{up}} \approx \Gamma_1^{\mathrm{low}} \approx 0.27$. Second, we demodulated white flux noise for both domains as a function of probe tone frequency $f_{\mathrm{exc}}$ at constant power $P_{\mathrm{exc}}\approx-70$~dBm for three different cases: 1) Boxcar window $w[n]$ without discarding, 2) Boxcar window and one period discarded and 3) Hamming window with one period discarded. As we mentioned earlier in section~\ref{subsubsec:frm}, discarding one period is sufficient to avoid the transient present at the beginning of each period of the sawtooth-shaped flux signal. For all cases we used $n_{\Phi_0}=4$ and the parameters in table \ref{tab:readoutparams}. The results  presented in figure \ref{fig:windows} for the non-discarding case were $\sqrt{S_{\Phi}^{\theta}} \approx 0.52 \mu\Phi_0/\sqrt{Hz}$ and $\sqrt{S_{\Phi}^{\gamma}} \approx 0.62 \mu\Phi_0/\sqrt{Hz}$ for phase and amplitude respectively, while in the second case $\sqrt{S_{\Phi}^{\theta}} \approx 0.61 \mu\Phi_0/\sqrt{Hz}$ and $\sqrt{S_{\Phi}^{\gamma}} \approx 0.72 \mu\Phi_0/\sqrt{Hz}$. Lastly, using Hamming window, $\sqrt{S_{\Phi}^{\theta}} \approx 0.71 \mu\Phi_0/\sqrt{Hz}$ and $\sqrt{S_{\Phi}^{\gamma}} \approx 0.84 \mu\Phi_0/\sqrt{Hz}$. These values are in well agreement with expressions \ref{eq:phase_tn_frm} and \ref{eq:gamma_tn_frm} using the aforementioned $\Theta_1$ and $\Gamma_1$ values. As expected, discarding a period yielded a degradation of $\sqrt{4/3} \approx 1.15$, while the Hamming window $\sqrt{4\kappa /3} \approx 1.35$. This corresponds to a factor $\kappa=1.37$ coincident with the equivalent noise bandwidth of the Hamming window with respect to the Boxcar. Although the Hamming window improves linearity\cite{SalumMuscheidFuster2023_1000162235}, its main-lobe is wider compared to the boxcar window leading to an increase in the noise level. For the purpose of this work, this degradation is not important since it affects both domains equally.

\begin{figure}
\includegraphics[width=0.48\textwidth]{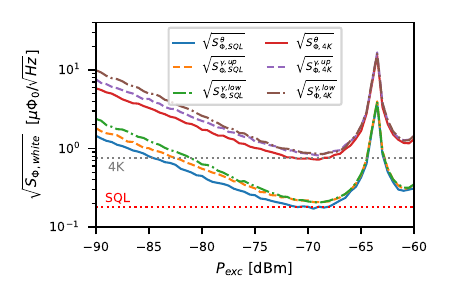}
\caption{\label{fig:frm-opt} Flux-ramp demodulated minimum white flux noise density $\sqrt{S_{\Phi,white}}$ for every demodulation domain as a function of power $P_{\mathrm{exc}}$ and additive noise temperature of $T_n=4$~K. Solid lines correspond to minimum noise for resonator’s phase readout while dashed and dashdotted lines are upper and lower local minimums for scattering amplitude readout. As a reference, dotted red and grey lines represent the minimum noise values for resonator's phase demodulation with a power of $P_{exc}=-70$~dBm in the case of the Standard Quantum Limit (SQL) and $T_n=4$~K respectively. Hamming window functions with $n_{disc}=1$ and $n_{\Phi_0}=4$ were used.}
\end{figure}

Similar to the open-loop demodulation, the noise in the optimal trajectories for both domains was determined for the SQL and $T_\mathrm{n}=4$~K. The results are shown in figure \ref{fig:frm-opt}. Again, resonator's phase readout provides the best performance and noise densities decreases with $\sqrt{S_{\Phi,white}} \propto 1/\sqrt{P_{\mathrm{exc}}}$ for both domains reaching minimums at $P_{exc}=-70$~dBm. Unlike the open-loop case, both amplitude minima differ at low powers, while at high powers they coincide due to the sinusoidal behaviour explained earlier. In agreement with equations \ref{eq:phase_tn_frm} and \ref{eq:gamma_tn_frm}, noise values in the SQL are $\sqrt{S_{\Phi,SQL}^{\theta}} \approx 0.18 \mu\Phi_0/\sqrt{Hz}$ and $\sqrt{S_{\Phi,SQL}^{\gamma}} \approx 0.22 \mu\Phi_0/\sqrt{Hz}$ being $\sqrt{T_{4K}/T_{SLQ}}\approx 4$ lower than the $T_n=4$~K case. As expected, these values represent a degradation of $c_{\mathrm{deg}}^{\theta} \approx 2.53$ and $c_{\mathrm{deg}}^{\gamma} \approx 2.8$ with respect to open-loop demodulation. Smaller values can be achieved removing the windows or not discarding periods depending on the linearity requirements or flux-ramp transient duration. Despite the imposed degradation, FRM provides other advantages as we will demonstrate below using our simulation framework in the following sections.

\subsubsection{\label{sec:frm-noiseequiv}TLS and Multiplicative Noise Impact for FRD}

\begin{figure}
\includegraphics[width=0.48\textwidth]{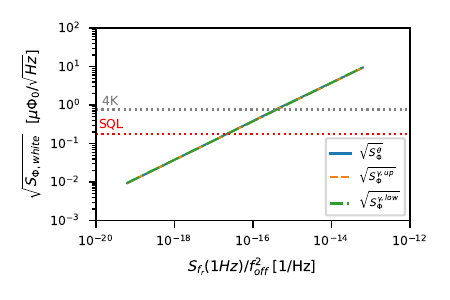}
\caption{\label{fig:frm-equivtls} Flux-ramp demodulated white flux noise density $\sqrt{S_{\Phi,white}}$ as a function of the Two-Level Systems (TLS) noise level of $S_{f_r}/f_{\mathrm{off}}$ for both demodulation domains in the minimum additive noise condition for a constant $P_{exc}=-70$~dBm. As a reference, dotted red and grey lines represent the minimum noise values for resonator's phase demodulation with a power of $P_{exc}=-70$~dBm in the case of the Standard Quantum Limit (SQL) and $T_n=4$~K respectively. Hamming window function with $n_{disc}=1$ and $n_{\Phi_0}=4$ were used ($\alpha=4/3$).}
\end{figure}

As shown in section \ref{sec:opl-multip}, we performed simulations to evaluate the impact of both TLS noise and multiplicative readout system noise for FRD. Starting with TLS, we determined the demodulated noise $\sqrt{S_{\Phi,white}}$ at the detector frequencies $f_{\mathrm{det}}$, for the same fractional frequency densities $S_{f_r}(1Hz)/f_{\mathrm{off}}^2$ values used before. The results are shown in figure \ref{fig:frm-equivtls} for both domains at the optimum conditions represented by stars in figure $\ref{fig:frm-noise}$. It can be seen that although the noise level increased with respect to the un-modulated case, for the same noise density TLS noise represents a lower impact with respect to additive noise levels represented by the red and grey dotted lines. This is due to demodulation process that down-converts the information signal from the carrier frequency $f_{\mathrm{mod}} \approx 61$~kHz, and therefore the noise evaluated at the same frequency $S_{f_r}(f_{\mathrm{mod}})/f_{\mathrm{off}}^2$. This value is considerably lower than the value at $S_{f_r}(f_{\mathrm{det}})/f_{\mathrm{off}}^2$ due to the $1/f$ TLS dependency. Therefore, relaxing the maximum TLS noise requirements to $S_{f_r}(1 Hz)/f_{\mathrm{off}}^2=2\cdot 10^{-17}$~Hz$^{-1}$ and $S_{f_r}(1 Hz)/f_{\mathrm{off}}^2=4\cdot 10^{-16}$~Hz$^{-1}$ for the SQL and $T_n=4$~K respectively. As far as the readout domain is concerned, results show again no preference for resonator's phase or scattering amplitude demodulation at higher powers, leaving multiplicative noise as the main factor to decide if there is an optimal readout domain.

\begin{figure}
\includegraphics[width=0.48\textwidth]{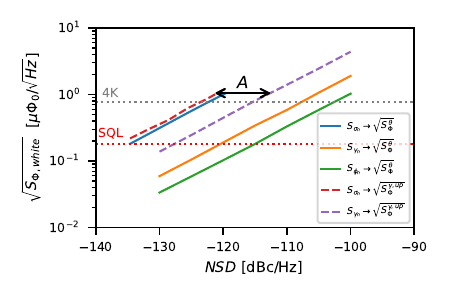}
\caption{\label{fig:frm-equiv} Flux-ramp demodulated white flux noise density for every demodulation domain and power for additive noise with $4K$. As a reference, dotted red and grey lines represent the minimum noise values for resonator's phase demodulation with a power of $P_{\mathrm{exc}}=-70$~dBm in the case of the Standard Quantum Limit (SQL) and $T_n=4$~K respectively. Hamming window functions with $n_{\mathrm{disc}}=1$ and $n_{\Phi_0}=4$ were used. As an example, black arrow represents the amplitude noise rejection factor $A$ in scattering amplitude domain readout.}
\end{figure}

Demodulated noise results for multiplicative sources under the optimum readout condition for additive noise found in section \ref{sec:frm-optprams} as a function of NSD are shown in figure \ref{fig:frm-equiv}. As seen in previous results, resonator's phase domain is the optimal domain for the additive noise case while the other noise sources remain negligible for the same NSD values. All noise demodulated values increase with respect to the open-loop case, following the same dependence with $\sqrt{S_{\Phi}} \propto \sqrt{NSD}$. However, all previously defined rejection factors decreased from $10$~dB to $8$~dB for amplitude noise in the scattering amplitude domain, while for the resonator's phase demodulation from $19$~dB to $18$~dB and from $40$~dB to $14$~dB for phase and amplitude noise respectively. As expected, all rejection factors decreased as a consequence of the FRM. Particularly, phase demodulation became more sensitive to amplitude noise when the readout tone is positioned further away from the resonance, where its projection is considerably greater and the gain is lower. On the contrary, phase noise in these positions is lower, therefore it has less impact. However, as we demonstrated before, system noise depends on the characteristics of the readout hardware used.

\subsubsection{\label{sec:sdr-example-frm}Optimum Domain for FRD}

\begin{figure}
\includegraphics[width=0.48\textwidth]{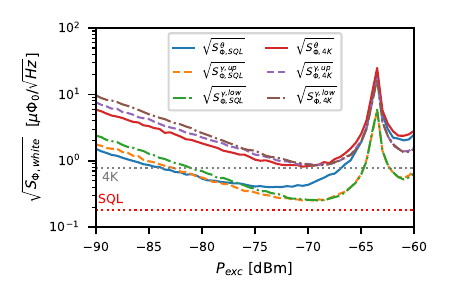}
\caption{\label{fig:readout-example-frm} Flux-ramp demodulated white flux noise density $\sqrt{S_{\Phi,white}}$ for every demodulation domain as a function of power $P_{\mathrm{exc}}$ and additive noise temperature of $T_n=4$~K. As a reference, dotted red and grey lines represent the minimum noise values for resonator's phase demodulation with a power of $P_{exc}=-70$~dBm in the case of the Standard Quantum Limit (SQL) and $T_n=4$~K respectively. Hamming window functions with $n_{\mathrm{disc}}=1$ and $n_{\Phi_0}=4$ were used.}
\end{figure}

Similar to the open-loop demodulation case, we used the simulation framework in order to predict the system noise performance of the SDR system used as an example. Measured noise values shown in figure \ref{fig:readout-noise} using magenta and cyan triangles were included in our simulation framework and again, TLS noise was not considered. The results of the flux-ramp demodulated noise for every domain and both SQL and $T_\mathrm{n}=4$~K additive noise conditions are shown in figure \ref{fig:readout-example-frm}. Once again, scattering amplitude readout shows the best performance when a real SDR readout system is used, and the impact of multiplicative noise became more relevant for the SQL scenario. In the case of the resonator's phase domain it is completely dominated by the phase noise of the readout system which has a value of $S_{\phi}(f_{\mathrm{mod}})\approx-108$~dBc/Hz at the frequency $f_{\mathrm{mod}}$. Amplitude noise at the same frequency with $S_{\gamma}(f_{\mathrm{mod}})\approx-130$~dBc/Hz does not have NSD to represent a considerable degradation for any of both scenarios. Clearly, like TLS, FRM helped on reducing the low-frequency multiplicative noise compared to the open-loop scheme. Quantitatively, noise levels at $P_{\mathrm{exc}}=-70$~dBm are $\sqrt{S_{\Phi}^{\theta}} \approx 0.43 \mu\Phi_0/\sqrt{Hz}$ and $\sqrt{S_{\Phi}^{\gamma}} \approx 0.24 \mu\Phi_0/\sqrt{Hz}$ for the SQL and $\sqrt{S_{\Phi}^{\theta}} \approx 0.84 \mu\Phi_0/\sqrt{Hz}$ and $\sqrt{S_{\Phi}^{\gamma}} \approx 0.90 \mu\Phi_0/\sqrt{Hz}$ for the $T_\mathrm{n}=4$~K case, all of them consistent with the values shown in figure \ref{fig:frm-equiv} for their respective NSD levels.

\section{\label{sec:disc}Discussion}

As a summary, table~\ref{tab:noisesumm} condenses all demodulated noise values for both demodulation schemes and readout domains. Here, $\sqrt{S_{\Phi}^{T_{n}}}$ represents the demodulated noise levels considering only the additive noise contribution $T_\mathrm{n}$, while $\sqrt{S_{\Phi}^{T_\mathrm{sys}}}$ takes into account the contributions of multiplicative and additive noise acting simultaneously. Based on these results and using equations \ref{eq:phase_tn},\ref{eq:amp_tn}, \ref{eq:phase_tn_frm} and \ref{eq:gamma_tn_frm}, it is possible to calculate an equivalent system temperature $\sqrt{S_{\Phi}^{T_\mathrm{sys}}}$ which reflects the measured noise. This would be the noise temperature that one would infer from the flux noise measurements and attribute it directly to the additive noise, not knowing that it is also produced by the multiplicative noise.

Table~\ref{tab:noisesumm} shows a large difference between $T_{n}$ and $T_{\mathrm{sys}}$ for the Open-loop demodulation of the resonator's phase in contrast to the scattering amplitude. This is a result of the excessive phase noise level measured at detector frequencies $S_{\phi}(f_{\mathrm{det}})$ compared to the additive $S_{\sigma}(f_{\mathrm{det}})$ and amplitude noise $S_{\gamma}(f_{\mathrm{det}})$ as can be seen in figure~\ref{fig:readout-noise}. For the case of FRD, the difference between $T_{n}$ and $T_{\mathrm{sys}}$ is lower due to the fact that additive and multiplicative spectral densities are evaluated at modulation frequencies $f_{\mathrm{mod}}$ (i.e. $S_{\sigma}(f_{\mathrm{mod}})$,$S_{\phi}(f_{\mathrm{mod}})$,$S_{\gamma}(f_{\mathrm{mod}})$) where there is not significant difference between noise levels. Therefore, the system noise temperature tends to the additive noise temperature $T_{\mathrm{sys}}\rightarrow T_{n}$. We can conclude that for this particular SDR system, the optimal readout domain is the scattering amplitude $\gamma$ and that the FRD allows to avoid the multiplicative noise at low frequencies representing a reduction in the demodulated flux noise. Although the multiplicative noise can be mitigated, it still represents a degradation in sensitivity particularly when working in conditions close to the SQL.

\begin{table}
\caption{\label{tab:noisesumm} Comparative between demodulated flux noise density $\sqrt{S_{\Phi}^{T_{n}}}$ considering only the additive noise contribution described with $T_{n}$ and the demodulated flux noise density $\sqrt{S_{\Phi}^{T_\mathrm{sys}}}$ considering all noise sources acting together as an equivalent additive system noise temperature $T_{\mathrm{sys}}$. The results are presented for both readout schemes and demodulation domains. In the Standard Quantum Limit (SQL) the additive noise temperature corresponds to $T_{n}\approx0.24$~K at $f_{\mathrm{exc}}=5$~GHz. Here, OPL stands for Open-Loop.}
\begin{ruledtabular}
\begin{tabular}{llcccc}
Scheme&Domain&$T_{n}$&$\sqrt{S_{\Phi}^{T_{n}}}$&$T_{\mathrm{sys}}$&$\sqrt{S_{\Phi}^{T_\mathrm{sys}}}$\\
\hline
OPL&$\theta$&$\approx 0.24$&$0.071\mu\Phi_0/\sqrt{Hz}$&29&$0.82\mu\Phi_0/\sqrt{Hz}$\\
OPL&$\theta$&4&$0.30\mu\Phi_0/\sqrt{Hz}$&33&$0.87\mu\Phi_0/\sqrt{Hz}$\\
OPL&$\gamma$&$\approx 0.24$&$0.08\mu\Phi_0/\sqrt{Hz}$&2&$0.24\mu\Phi_0/\sqrt{Hz}$\\
OPL&$\gamma$&4&$0.34\mu\Phi_0/\sqrt{Hz}$&5.6&$0.40\mu\Phi_0/\sqrt{Hz}$\\
FRD&$\theta$&$\approx 0.24$&$0.18\mu\Phi_0/\sqrt{Hz}$&1.4&$0.43\mu\Phi_0/\sqrt{Hz}$\\
FRD&$\theta$&4&$0.71\mu\Phi_0/\sqrt{Hz}$&5.1&$0.84\mu\Phi_0/\sqrt{Hz}$\\
FRD&$\gamma$&$\approx 0.24$&$0.22\mu\Phi_0/\sqrt{Hz}$&0.4&$0.24\mu\Phi_0/\sqrt{Hz}$\\
FRD&$\gamma$&4&$0.84\mu\Phi_0/\sqrt{Hz}$&4.5&$0.90\mu\Phi_0/\sqrt{Hz}$\\
\end{tabular}
\end{ruledtabular}
\end{table}

Simulations presented in this article contributed to improve system noise estimations finding the maximum noise values that can be managed for a given sensitivity. Thus, it enables the optimization of the associated \textmu MUX, cold and warm-temperature readout systems. Although for this work the criterion for the selection of optimal parameters was found to be based on the condition of minimum readout noise, two important factors must be taken into account. First, for a large number of channels $N$, the total power at the output of the \textmu MUX grows with $P_{\mathrm{out}} \propto N$ and can saturate the HEMT amplifier degrading the noise performance\cite{Rao2020}. Particularly in the case presented in \ref{sec:frm-optprams} there is no substantial difference, the resonator's phase demodulation presented $P_{\mathrm{out}}\approx -82$~dBm with respect to the scattering amplitude with  $P_{\mathrm{out}}\approx -79$~dBm for a $P_{\mathrm{exc}}=-70$~dBm. Second, while the phase demodulation yielded better results for the additive noise case, the scattering amplitude is robust to other types of noise, as well as not requiring a resonance circle transformation which can be affected by phase variations\cite{Silva2022} of the RF components and requires more digital resources to be implemented\cite{Gard2018,Karcher2022}.

As mentioned at the beginning of this article, the presented analysis applies only to readout by a single fixed tone. A recently proposed method called Tone-Tracking tries to overcome the aforementioned limitation imposed by the saturation of the HEMT amplifier when reading a large number of detectors. This system uses a feedback system that corrects the frequency of the probing tone to keep it in resonance, and consequently minimizing the peak power at the \textmu MUX output\cite{Yu2023,yusim}. Given the advantages of this system, future work will consider extending the results of this work to the analysis of this innovative readout technique.

\section{\label{sec:conc}Conclusion}

We successfully extended the capabilities of previously developed simulation frameworks for \textmu MUX readout including multiplicative noise sources, as well as the ability of demodulating data in different domains. Through different simulations the optimum readout parameters to achieve the lowest possible readout noise for a \textmu MUX device optimized for bolometric applications were found in both open-loop and flux-ramp demodulation schemes. We probe that the optimal readout parameters in the case of open-loop and flux-ramp demodulation are different and mainly determined by the first harmonic component used for demodulation. We demonstrate that the dominant noise source in both cases is the additive noise and the optimal demodulation domain for this condition is the resonator's phase. Under the minimum noise parameters for additive noise, the impact of the multiplicative sources was assessed. An example using a typical SDR readout system was presented and the impact on the system noise estimated. Due to the higher readout phase noise, the results showed considerable degradation of the system noise when demodulation is performed in the resonator's phase domain. Contrarily, demodulation in scattering amplitude domain yielded the minimum noise dominated by additive noise. As expected, degradation becomes more evident close to the SQL when parametric amplifiers are used. Additionally and as an integral part of this work, the performance of the flux-ramp demodulation using different windows, the effect of discarding periods, as well as the demodulation of different harmonic components was evaluated and contrasted with the open-loop case. Last but not least, it is important to note that the noise analysis of the readout system presented here is not only useful for the design of SDR readout for \textmu MUX systems, but also for different frequency domain multiplexed superconducting devices such as MKIDs and QUBITs.

\section*{Data Availability Statement}

The data and artwork that support the findings of this study are available from the corresponding author upon reasonable request.

\section*{Acknowledgments}
M. E. García Redondo is supported by the Comisión Nacional de Energía Atómica (CNEA) as well as for the Helmholtz International Research School in Astroparticles and Enabling Technologies (HIRSAP). M. E. García Redondo also acknowledges the support of the Karlsruhe School of Elementary and Astroparticle Physics: Science and Technology (KSETA).

\appendix

\section{\label{app:readnoisecharact}Readout System Noise Characterization}

\begin{figure}
\includegraphics[width=0.45\textwidth]{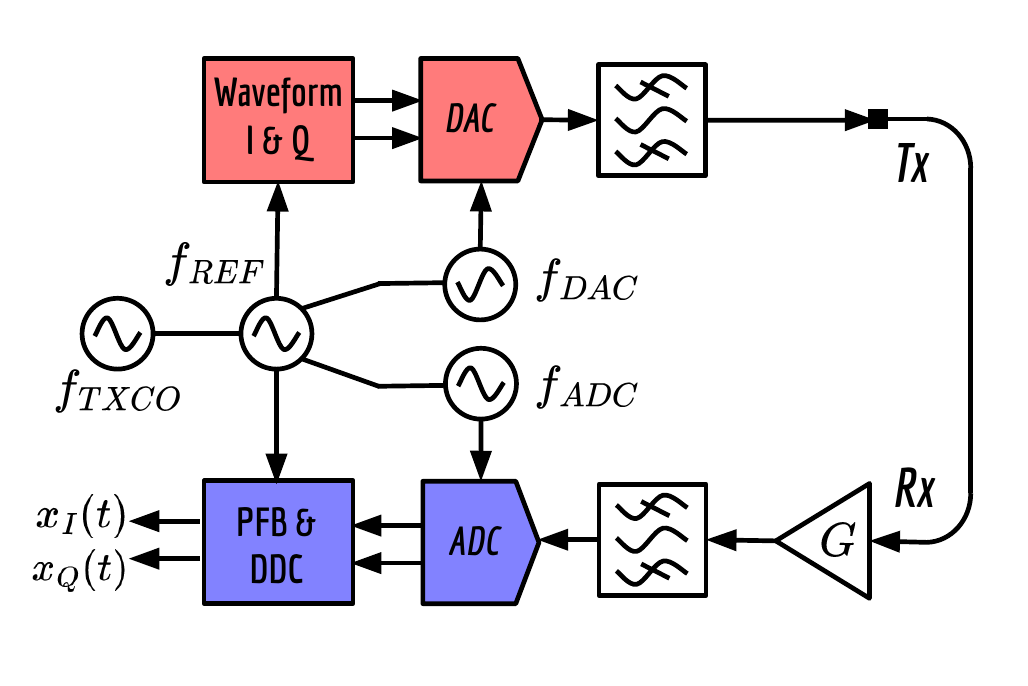}
\caption{\label{fig:rfsocscheme} Noise characterization set-up for the Direct-RF Software-Defined Radio (SDR) readout system based on the RFSoC ZCU216 evaluation kit\cite{zcu216}. A single tone at $f_{\mathrm{exc}}=5$~GHz is directly generated with the DAC sampling at $f_{DAC}=8$~GHz and filtered by a band-pass filter. The tone at the transmitter (Tx) is sent directly to the receiver port (Rx) where it is amplified and filtered again. Later, an ADC samples the signal in a higher order Nyquist zone with a sample rate equal to $f_{ADC}=2$~GHz. Finally, the tone is channelized by a Poly-Phase Filter Bank (PFB) and Digitally Down-Converted (DDC) to baseband.}
\end{figure}

The measurement set-up for the SDR readout noise characterization is shown in figure \ref{fig:rfsocscheme}. It comprises of a Direct-RF Software-Defined Radio (SDR) readout system based on the RFSoC ZCU216 evaluation kit\cite{zcu216}. In this case, a single tone at frequency $f_{\mathrm{exc}}=5$~GHz is directly generated in the 2\textsuperscript{nd} Nyquist zone using a high-speed Digital-to-Analog Converter (DAC) sampling at $f_{\mathrm{DAC}}=8$~GHz. The tone is filtered by a bandpass filter in order to eliminate spurious components and sent to the receiver (Tx to Rx loopback). At the receiver side,  the tone is amplified by a commercial LNA and filtered again. After that, it is acquired by an Analog-to-Digital Converter (ADC) in the 6\textsuperscript{th} Nyquist zone. Finally, it is channelized in the digital domain by a Poly-phase Filter Bank (PFB), Down-converted to baseband and decimated using a Digital-Down Converter (DDC). The cutoff frequency of the low-pass filter used in the DDC is $f_{\mathrm{cut}}\approx1.6$~MHz. At this point, $x_I(t)$ and $x_Q(t)$ components are available at a data rate of $f_s \approx 15.62$~MHz. This system is equivalent to the homodyne system presented in figure~\ref{fig:readout}. The frequency distribution is performed through CLK104 Add-on Board~\cite{clk104} (see figure~\ref{fig:rfsocscheme}). On it, a Temperature Compensated Crystal Oscillator (TXCO) generates a $10$~MHz signal that feed a dual loop jitter cleaner and clock distribution chip (LMK04828B). This chip is responsible for generating clock signals for the digital system as well as two $250$~MHz reference signals for the DAC and ADC sampling clock synthesizers. Two different RF synthesizers (LMX2594) connected to the RFSoC chip generate the $f_{\mathrm{DAC}}=8$~GHz and $f_{\mathrm{ADC}}=2$~GHz sampling clocks from the $250$~MHz references. We are aware that this configuration might not be the most suitable in terms of phase noise compared to traditional SDR systems using RF mixing boards~\cite{Gard2018,Rantwijk,Yu2023,Redondo2023rfsoc}. Nevertheless, it is a good example in order to demonstrate the importance of selecting a suitable frequency reference and clock generation system. Future works include the optimization of the PLLs and the use of an external ultra-low phase noise reference (e.g. Rubidium oscillator).

Using the probing tone complex envelope $x_I(t)+jx_Q(t)$, phase $\phi(t)=\arctan{(x_Q(t)/x_I(t))}$ and amplitude $\gamma(t)=|x_I(t)+jx_Q(t)|$ time traces were computed. By means of Python, $N_{s}=2^{24}$ samples were used to calculate readout system phase $S_{\phi}^{\mathrm{meas}}(f)$ and amplitude $S_{\gamma}^{\mathrm{meas}}(f)$ noise spectra. In order to avoid spectral leakage, a Blackman–Harris windows was applied achieving a spectral resolution of around $\approx 1$~Hz. The resulting spectra for an ADC input power of $P_{ADC}\approx-5$~dBm ($P_{Tx}\approx-25$~dBm) are shown in figure~\ref{fig:readout-noise}. Equivalently to equation~\ref{eq:signalmeas} with $S_{21}\approx1$ (i.e. in loopback), additive and multiplicative noise are measured simultaneously and cannot be distinguished. This is because the additive noise (orange noise cloud in figure~\ref{fig:trajectories}) can also be interpreted as simultaneous variations in phase and amplitude. In this case, the phase $S_{\phi}^{\mathrm{meas}}(f)$ and amplitude $S_{\gamma}^{\mathrm{meas}}(f)$ measured spectra consist of the two contributions and can be expressed as,

\begin{equation}
    S_{\phi}^{\mathrm{meas}}(f)= S_{\phi}(f) + \frac{k_B T_{n}^{ADC}}{P_{ADC}}
    \label{eq:measphase}
\end{equation}

\begin{equation}
    S_{\gamma}^{\mathrm{meas}}(f)= S_{\gamma}(f) + \frac{k_B T_{n}^{ADC}}{P_{ADC}}
    \label{eq:measamp}
\end{equation}

\begin{figure}
\includegraphics[width=0.45\textwidth]{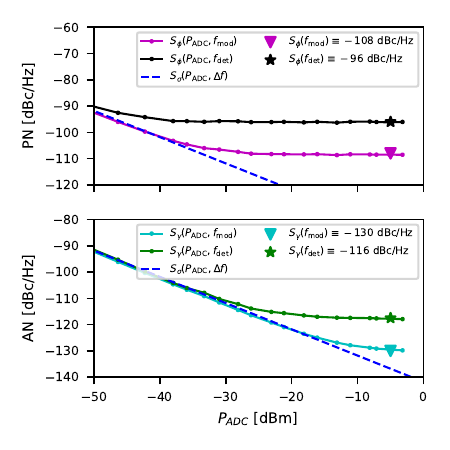} 
\caption{\label{fig:noisepwsweep} Noise spectral densities of a tone at $f_{\mathrm{exc}}=5$~GHz as a function ADC input power $P_{ADC}$ measured at frequencies $f_{\mathrm{det}}$ and $f_{\mathrm{mod}}$. Top) Phase spectral density $S_{\phi}(P_{\mathrm{P_{ADC}}})$ . Bottom) Amplitude spectral density measured at frequencies $f_{\mathrm{det}}$ and $f_{\mathrm{mod}}$. Top) Phase spectral density $S_{\phi}(P_{\mathrm{P_{ADC}}})$. Dashed blue straight line represents the additive noise contribution while dashed black, green stars and magenta, cyan triangles corresponds to the phase $S_{\phi}$ and amplitude $S_{\gamma}$ multiplicative noise values adopted for the analysis and represented in figure~\ref{fig:readout-noise}. These values were measured at an ADC input power of $P_{ADC} \approx -5$~dBm.}
\end{figure}

Here, $P_{ADC}$ is the ADC input power and $T_{n}^{ADC}$ is the ADC input referred additive noise temperature. The terms in the left side of equations~\ref{eq:measphase} and~\ref{eq:measamp} are the pure multiplicative spectra while the additives are in the right side. In order to extract the pure multiplicative phase and amplitude noise level, a power sweep was performed until the dependence on power vanishes (i.e. $S_{\phi}^{meas}(f)\approx S_{\phi}(f)$ and $S_{\gamma}^{meas}(f) \approx S_{\gamma}(f)$). The power sweep was performed modifying the DAC output power $P_{DAC}$. The result of these measurements are shown in figure~\ref{fig:noisepwsweep}. At the top panel, solid black and magenta lines represent the phase noise measured at detector frequencies $S_{\phi}(P_{ADC},f_{\mathrm{det}})$ and modulation frequencies $S_{\phi}(P_{ADC},f_{\mathrm{mod}})$ respectively. The bottom panel is equivalent, but for the amplitude noise case showing $S_{\gamma}(P_{ADC},f_{\mathrm{det}})$ and $S_{\gamma}(P_{ADC},f_{\mathrm{mod}})$.

At lower powers, both phase and amplitude spectra are dominated by the additive noise following the trajectory indicated with the straight dashed blue line. On the contrary, at high powers phase and amplitude noise densities are dominated by the multiplicative noise and reach a constant value. As a result, the true phase $S_{\phi}$ and amplitude $S_{\gamma}$ multiplicative noise values adopted for the analysis described in the main text are $S_{\phi}(f_{\mathrm{det}})=-96$~dBc/Hz, $S_{\phi}(f_{\mathrm{mod}})=-108$~dBc/Hz, $S_{\gamma}(f_{\mathrm{det}})=-116$~dBc/Hz and $S_{\gamma}(f_{\mathrm{mod}})=-130$~dBc/Hz. These values are represented by black and green stars, and magenta and cyan triangles in figure~\ref{fig:noisepwsweep} and in the manuscript body in figure~\ref{fig:readout-noise}. Additionally, this procedure allows us to determine the additive noise level from the linear tendency given by the blue dashed line. Using equation~\ref{eq:addnsd} and considering $S_{\sigma}(\Delta f)\approx -135$~dBc/Hz at an ADC power of $P_{ADC}\approx-5$~dBm, we calculated the ADC input referred additive noise temperature yielding to $T_{n}^{ADC} \approx 724 \cdot 10^3$~K. This temperature cannot be directly compared to the cryogenic LNA equivalent noise temperature, therefore, it should be referred to the \textmu MUX output by dividing the noise temperature $T_{n}^{ADC}$ by the gain $G_{Rx}$ measured between the \textmu MUX output and the ADC input (i.e. $T_{n}=T_{n}^{ADC}/G_{Rx}$). Since $G_{Rx}$ is an unknown value and considering that the SDR system was designed properly, we will consider that the SDR system only contributes to multiplicative noise and that the additive noise temperature $T_n$ is dominated by the cryogenic amplifier (e.g. the HEMT or the TWPA~\cite{Malnou,Zobrist19}).

\section{\label{app:synthpn}Synthesizer Phase Noise}

\begin{figure}
\includegraphics[width=0.45\textwidth]{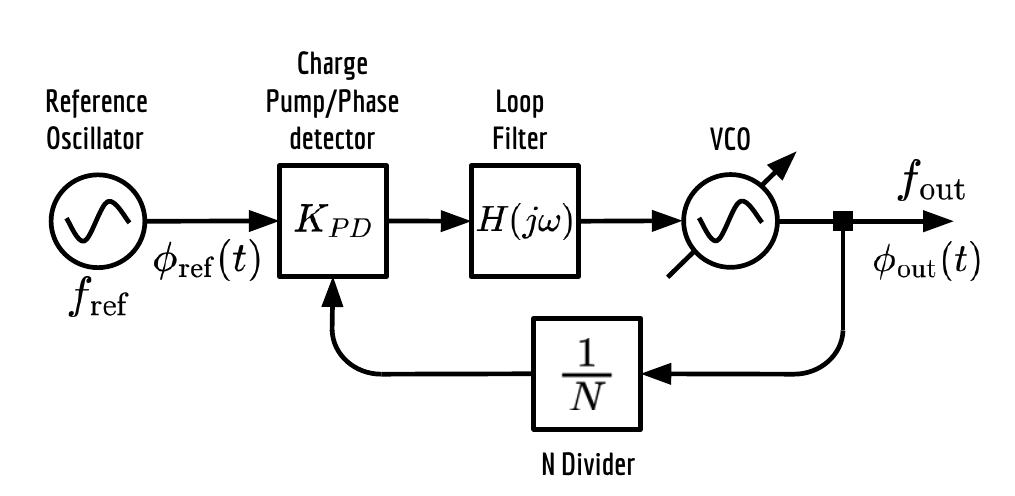}
\caption{\label{fig:plldiag} Phase-Locked Loop (PLL) block diagram. The synthesizer creates a frequency $f_{\mathrm{out}}$ from a reference oscillator of $N$ times lower frequency $f_{\mathrm{ref}}$. The feedback system allows the output phase $\phi_{\mathrm{out}}$ to be locked to the reference phase $\phi_{\mathrm{ref}}(t)$. The phase noise response is determined by the noise contributed by each block and the closed-loop gain between each point in the system and the output $\phi_{\mathrm{out}}(t)$.}
\end{figure}

The CLK104 Add-on Board is equipped with two wideband Phase-Locked Loop (PLL) synthesizers (LMX2594) used to generate DAC and ADC sampling clocks (see figure~\ref{fig:rfsocscheme}). Most of the phase noise characteristics of the sampling clocks are determined by the PLL components. A simplified block diagram of a PLL is shown in figure \ref{fig:plldiag}. It comprises five blocks: 1) a Reference Oscillator, 2) a Voltage-Controlled Oscillator (VCO), 3) a frequency $N$ Divider. 4) a Loop-Filter, and 5) a Phase Detector.

When the system is locked, the VCO generates an output frequency $f_{\mathrm{out}}$ which is divided $N$ times and sent to the phase detector. The phase detector compares the phases of the signal coming from the divider $\phi_{\mathrm{out}}(t)/N$ with that of the reference oscillator $\phi_{\mathrm{ref}}(t)$. At its output, the phase detector (usually implemented as a charge pump) generates a voltage proportional to the phase difference $V_{PD}(t)=K_{PD}(\phi_{\mathrm{ref}}(t)-\phi_{\mathrm{out}}(t)/N$). This signal is filtered and used to close the loop by adjusting the VCO voltage $V_{VCO}(t)$ such that the generated frequency is in phase with that of the reference oscillator. In this condition we can state that the output signal is an exact multiple of the frequency of the reference oscillator $f_{\mathrm{out}}=Nf_{\mathrm{ref}}$ and follows its phase variations. Typically the control loop is designed so that the phase error in the steady state is zero (i.e. $\phi_{\mathrm{ref}}(t)-\phi_{\mathrm{out}}(t)/N=0$).

\begin{figure}
\includegraphics[width=0.45\textwidth]{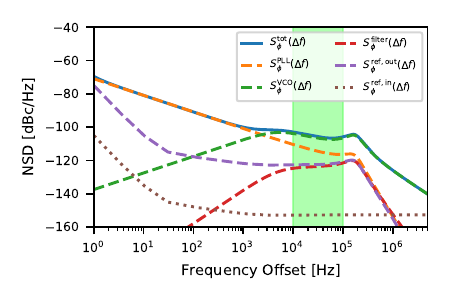}
\caption{\label{fig:pllcontrib} Phase-Locked Loop phase noise contributions for a frequency $f_{\mathrm{out}}=8$~GHz generated from a reference frequency of $f_{\mathrm{ref}}=250$~MHz. The solid blue line represents the total phase noise $S_{\phi}^{tot}(\Delta f)$, while dashed lines are the phase noise contributions $S_{\phi}^{i}(\Delta f)$ of each block shown in figure~\ref{fig:plldiag}. Dotted brown line corresponds to the reference oscillator phase noise $S_{\phi}^{\mathrm{ref,in}}(\Delta f)$ at the input, while $S_{\phi}^{\mathrm{ref,out}}(\Delta f)$ is the oscillator phase noise referred to the output.}
\end{figure}

The total phase noise $S_{\phi}^{tot}(\Delta f)$ at the PLL output is determined by the contribution of each of the PLL components. Using the closed-loop gain between each system component and the output $G_{i}(f)$, it is possible to calculate the total phase noise by adding all noise components $S_{\phi}^{i}(\Delta f)$. Using the simulation software provided by the manufacturer, the phase noise of each component of the LMX2594 synthesizer was estimated for the configuration used in the CLK104\cite{PLLsim,clk104}. For demonstration purposes, a $f_{\mathrm{ref}}=250$~MHz reference oscillator with phase noise characteristics $S_{\phi}^{\mathrm{ref,in}}(\Delta f)$ equivalent to a Rubidium oscillator was adopted for the simulations~\cite{FS725}. The simulation results for a frequency of $f_{\mathrm{out}}=8$~GHz are shown in figure~\ref{fig:pllcontrib}. In this case the synthesizer parameters were optimized to obtain the minimum phase noise in the green shaded region. Only software configurable parameters were modified, avoiding to modify the CLK104 hardware.

The results show that the total phase noise $S_{\phi}^{\mathrm{tot}}(\Delta f)$ for this configuration is determined by the internal components of the synthesizer. At low frequencies, it is dominated by the PLL (Divider and Phase Detector) noise $S_{\phi}^{PLL}(\Delta f)$ (dashed orange line), while at higher frequencies it is dominated by the VCO phase noise $S_{\phi}^{VCO}(\Delta f)$ (dashed green line). Furthermore, we can conclude that even with the use of a Rubidium oscillator the phase noise is dominated by the PLL circuitry. This is mainly due to three factors: First, the integrated VCO has high noise level at the operating frequency. Second, the high output frequency imposes a division factor $N=32$ amplifying both the divider and reference noise by $20\log_{10}(N)$ (see magenta line in the figure). Third, the PLL generates an electrical noise at the input of the VCO which creates a strong $1/f$ component given the integrating characteristics of the VCO~\cite{HajiPLL}. Therefore, the phase noise can only be improved by using a higher quality PLL and VCO as well as reducing the value of $N$ by using a higher frequency reference.

Phase noise present in the ADC or DAC sampling clocks is directly transferred to the generated/acquired tone. In this case the phase noise is dominated by the DAC sampling at $f_{\mathrm{out}}=8$~GHz. This is consistent with the measurement results shown in figure \ref{fig:readout-noise} and with experimental results shown in other articles related with Direct-RF SDR systems based on RFSoC devices used in the field of Quantum Computing~\cite{qick_sdr,exp_quick}. More details about phase noise in PLLs and its optimization can be found in the following references~\cite{HajimPN,HajiPLL}.

\section{\label{app:projections}Noise Projections}

\begin{figure}
\includegraphics[width=0.45\textwidth]{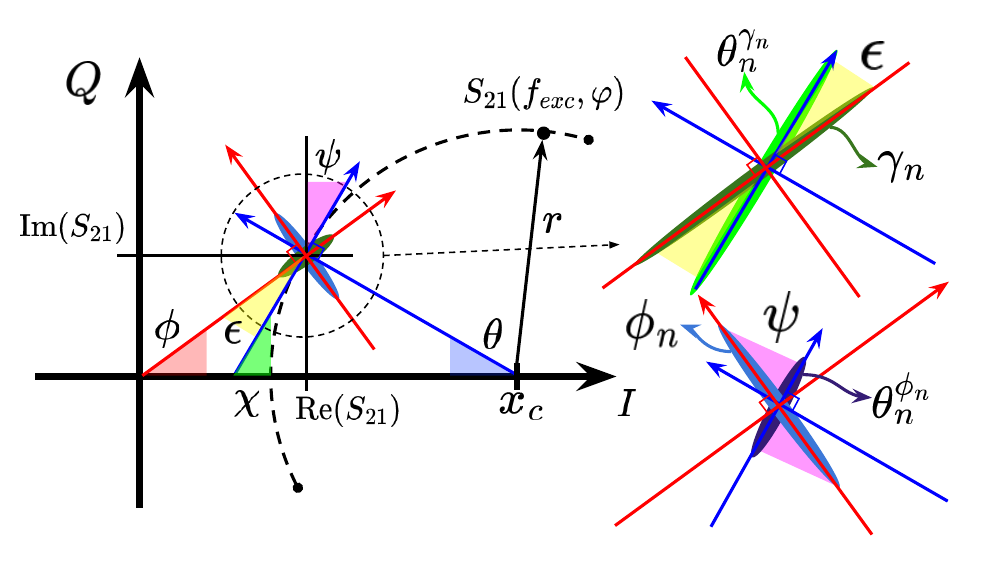} 
\caption{\label{fig:projection} Detailed description of the projections of each noise source for both demodulation domains. Amplitude of transmission scattering parameter $\gamma$ and resonator's phase $\theta$ measured form the resonator rotating frame.}
\end{figure}

Considering that noise amplitudes are considerably small respect to $S_{21}(t)$ variations, each noise source can be geometrically projected into the corresponding domain. Figure \ref{fig:projection} shows a detailed description of figure \ref{fig:trajectories}. Defining the readout and resonator's phases as,

\begin{equation}
    \phi(f_{\mathrm{exc}},\varphi)=\arctan{\left\{\frac{\operatorname{Im}[S_{21}(f_{\mathrm{exc}},\varphi)]}{\operatorname{Re}[S_{21}(f_{\mathrm{exc}},\varphi)]}\right\}}
\end{equation}

\begin{equation}
    \theta(f_{\mathrm{exc}},\varphi)=\arctan{\left\{\frac{\operatorname{Im}[S_{21}(f_{\mathrm{exc}},\varphi)]}{x_c-\operatorname{Re}[S_{21}(f_{\mathrm{exc}},\varphi)]}\right\}}
\end{equation}

Using triangles proprieties we easily demonstrate that $\psi=\pi/2-\epsilon=\theta+\phi$, Therefore, using the adequate trigonometric relations we can write the projections of amplitude $\gamma_n(t)$ and phase $\phi_n(t)$ readout noise into the resonator's phase $\theta(t)$ as follows,

\begin{equation}
    \theta_n^{\gamma_n}(t) \approx \gamma_n(t) \cos(\epsilon)=\gamma_n(t) \frac{\gamma}{r} \sin(\theta+\phi)
\end{equation}

\begin{equation}
    \theta_n^{\phi_n}(t) \approx \phi_n(t) \cos(\psi)=\phi_n(t)  \frac{\gamma}{r} \cos(\theta+\phi)
\end{equation}

Where the approximation $\theta \approx \Delta_{\perp}/r$ was used to convert absolute voltage variations into phase units. $r$ stands for the resonance circle radius, while $\Delta_{\perp}$ represents voltage variations perpendicular to $r$ or, equivalently, tangent to the resonance circle. The noise spectral densities in table \ref{tab:projectios} were calculated considering that, for a given scaling factor $g$ applied to signal $y(t)$, the resulting power spectral density of the scaled signal $z(t)$ is equal to $S_z(\Delta f)= g^2 S_y(\Delta f)$. 

\nocite{*}
\bibliography{aipsamp}

\end{document}